\documentclass[12pt,fleqn]{article}
\usepackage{color,amsmath,epsfig}

\newcommand{\be}{\begin{eqnarray}}
\newcommand{\ee}{\end{eqnarray}}
\newcommand{\nee}{\nonumber\end{eqnarray}}
\newcommand{\nn}{\nonumber}
\newcommand{\noi}{\noindent}

\newcommand{\mbf}      {\boldmath}
\newcommand{\sfrac}[2] {{\textstyle \frac{#1}{#2}}}
\newcommand{\smaf}[2]  {{\textstyle \frac{#1}{#2} }}

\newcommand{\eq}[1]  {\mbox{(\ref{eq:#1})}}
\newcommand{\fig}[1] {\mbox{Fig.~\ref{fig:#1}}}
\newcommand{\Fig}[1] {\mbox{Figure~\ref{fig:#1}}}

\def\a               {\alpha}
\def\b               {\beta}
\def\d               {\delta}
\def\g               {\gamma}
\def\G               {\Gamma}

\def\t               {\theta}

\def\x               {\chi}

\def\ti              {\tilde}

\def\sq              {\ti q}
\def\st              {\ti t}
\def\sb              {\ti b}
\def\ch              {\ti \x^\pm}
\def\chp             {\ti \x^+}
\def\chm             {\ti \x^-}
\def\nt              {\ti \x^0}
\def\sg              {\ti g}

\newcommand{\mst}[1]   {m_{\st_{#1}}}
\newcommand{\msb}[1]   {m_{\sb_{#1}}}
\newcommand{\mch}[1]   {m_{\ti \x^+_{#1}}}
\newcommand{\mnt}[1]   {m_{\ti \x^0_{#1}}}
\newcommand{\mhp}      {m_{H^+}}
\newcommand{\msg}      {m_{\ti g}}

\def\M               {{\cal M}}
\def\H               {{\cal H}}

\def\PL              {P_L^{}}
\def\PR              {P_R^{}}
\def\Rst             {R^{\,\st}}
\def\Rsb             {R^{\,\sb}}
\def\Rsq             {R^{\,\sq}}
\def\Rsts            {R^{\,\st *}}
\def\Rsbs            {R^{\,\sb *}}

\def\rzw             {\sqrt{2}}
\def\delr            {\!\stackrel{\leftrightarrow}{\partial^\mu}\!}

\newcommand{\gsim}{\;\raisebox{-0.9ex}
           {$\textstyle\stackrel{\textstyle >}{\sim}$}\;}

\renewcommand{\Re}{{\rm Re}}
\renewcommand{\Im}{{\rm Im}}

\def\lrd{\stackrel{\leftrightarrow}{\partial^{\mu}}}

\begin{document}


\begin{titlepage}
\begin{flushright}
   \vspace*{-2.4cm} CERN-TH/2002-106\\[-1mm]
                    HEPHY-PUB 756/02\\[-1mm]
                    hep-ph/0205227
\end{flushright}

\begin{center}

\vspace*{4mm}

{\large\bf\boldmath
    CP violation in charged Higgs boson decays \\
    in the MSSM with complex parameters}\\

\vspace{1.6cm}

{\bf Ekaterina Christova}\,\footnote{e-mail: echristo@inrne.bas.bg}\\
{\em Institute of Nuclear Research and Nuclear Energy, \\
     Tzarigradsko Chaussee 72, Sofia 1784, Bulgaria} \\

\vspace{0.5cm}

{\bf Helmut Eberl\,\footnote{e-mail: helmut@hephy.oeaw.ac.at},
     Walter Majerotto\,\footnote{e-mail: majer@hephy.oeaw.ac.at}} \\
{\em Institut f\"ur Hochenergiephysik der
     \"Osterreichischen Akademie der Wissenschaften, \\
     A-1050 Vienna, Austria}

\vspace{0.5cm}

{\bf Sabine Kraml}\,\footnote{e-mail: sabine.kraml@cern.ch}\\
{\em Theory Division, CERN, CH-1211 Geneva 23, Switzerland}

\vspace{0.5cm}

\end{center}

\vspace{0.5cm}
\begin{abstract}
Supersymmetric loop contributions can lead to different
decay rates of $H^+\to t\bar b$ and $H^-\to b\bar t$.
We calculate the decay rate asymmetry
$\d^{CP} = [\G(H^+\to t\bar b)-\G(H^-\to b\bar t)]\,/\,
           [\G(H^+\to t\bar b)+\G(H^-\to b\bar t)]$
at next-to-leading order in the MSSM with complex parameters.
We analyse the parameter dependence
of $\d^{CP}$ with emphasis on the phases of $A_t$ and $A_b$.
It turns out that the most important contribution comes from the
loop with stop, sbottom, and gluino. If this contribution is present,
$\d^{CP}$ can go up to $10-15\%$ for $\tan\b\sim 10$,
and to $\sim 5\%$ for large values of $\tan\b$.
\end{abstract}

\vfill

\end{titlepage}
\newpage
\setcounter{page}{1}

\section{Introduction}

Already for some time, it has been customary to look for CP-violating
effects beyond the Standard Model (SM). All extensions of the SM
contain possible new sources of CP violation through additional 
CP-violating phases. In particular, in the Minimal Supersymmetric Standard 
Model (MSSM), the Higgs mixing parameter $\mu$ in the superpotential,
two of the soft SUSY-breaking Majorana gaugino masses $M_i$ ($i=1,2,3$),
and the trilinear couplings $A_f$ (corresponding to a fermion $f$)
can have physical phases, which cannot be rotated away
without introducing phases in other couplings~\cite{Dugan:1984qf}.
From the point of view of baryogenesis, one might hope that these
phases are large~\cite{Carena:1997ki}.
On the other hand, the experimental limits on electron and neutron 
electric dipole moments (EDMs)~\cite{Altarev:cf},
$|d^e| \le 2.15\times 10^{-13}$~e/GeV,
$|d^n| \le 5.5\times 10^{-12}$~e/GeV, place severe constraints
on the phase of $\mu$, $\phi_\mu < {\cal O}(10^{-2})$~\cite{Nath:dn},
for a typical SUSY mass scale of the order of a few hundred GeV.
A larger $\phi_\mu$ imposes fine-tuned relationships between this
phase and other SUSY parameters~\cite{Ibrahim:1997nc}.
CP-violating effects that might arise from $A_{u,d}$ (where $u,d$
are light quarks) are very much suppressed as they are proportional
to $m_{u,d}$. On the other hand, the trilinear couplings of the
third generation $A_{t,b,\tau}$ can lead to significant CP-violation
effects, especially in top quark physics~\cite{Atwood:2000tu}.
Phases of $\mu$ and $A_{t,b,\tau}$ also affect the Higgs sector
in a relevant way. Although the Higgs potential of the MSSM is
invariant under CP at tree level, at loop level CP is sizeably
violated by complex couplings~\cite{Pilaftsis:1998pe,CEPW00}.
As a consequence, the three neutral Higgs mass eigenstates
are superpositions of the CP eigenstates $h^0$, $H^0$, and $A^0$.

In this paper, we study CP violation in the decays $H^+\to t\bar b$
and $H^-\to b\bar t$ in the MSSM with complex parameters.
In particular, we calculate the CP-violating asymmetry
\begin{equation}
  \d^{CP} = \frac{\G\,(H^+\to t\bar b)-\G\,(H^-\to b\bar t)}
                 {\G\,(H^+\to t\bar b)+\G\,(H^-\to b\bar t)}\,,
\label{eq:defDCP}
\end{equation}
due to one-loop exchanges of $\st$, $\sb$, $\sg$, $\ch$, $\nt$,
and $H^0$, see Fig.~1, taking into account CP violation in the
neutral Higgs system according to \cite{CEPW00}.
Of course, the diagrams of Fig.~1 only contribute to $\d^{CP}$
if they have an absorptive part.
Since $\phi_\mu$ is highly constrained, the most important phases in
our analysis are $\phi_t$ and $\phi_b$, the phases of $A_t$ and $A_b$.
Therefore, we expect the graph with $\st$, $\sb$, and $\sg$ in the
loop to be the most important one, and $\d^{CP}$ to be large
in the case $\mhp>\mst{1}+\msb{1}$. 
In principle, there would also be a contribution due to 
$\ti\nu$ and $\ti\tau$ exchange analogous to Fig.~1e. 
However, this can be neglected in our study. 

The paper is organized as follows:
In Section~2 we give the basic formulae for the $H^\pm\to tb$ 
decays and define the decay rate asymmetry $\d^{CP}$ 
at the 1-loop level in terms of CP-violationg form factors 
$\d Y_i^{CP}$ ($i=t,b$).
The explicit expressions for $\d Y_{t,b}^{CP}$ due to the diagrams 
of Fig.~1 are given in Section~3.
In Section~4, we perform a detailed numerical analysis.
In Section~5, we summarize our results and comment on the feasibility 
of measuring the CP-violating asymmetry $\d^{CP}$.
Appendices A, B, and C contain the necessary mass and mixing matrices,
the couplings, and the definition of the two- and three-point
functions used in this paper.

\section{The \boldmath $H^\pm$ decay}

The matrix elements of the $H^+\to t\bar b$ and $H^-\to b\bar t$ 
decays can be written as
\begin{eqnarray}
  {\cal M}_{H^+} &=& \bar u(p_t)\,\big[Y_b^+\PR+Y_t^+\PL\big]\,v(-p_b)\,,
    \label{eq:MtreeHp} \\
  {\cal M}_{H^-} &=& \bar u(p_b)\,\big[ Y_t^-\PR+Y_b^-\PL\big]\,v(-p_t)\,,
    \label{eq:MtreeHm}
\end{eqnarray}
with $P_{R,L}^{} = \frac{1}{2}(1\pm\g_5)$ and the loop-corrected couplings
\begin{equation}
  Y_i^{\pm} = y_i^{} + \d Y_i^{\pm} \quad (i=t,b) \,;
  \label{eq:Yi}
\end{equation}
$y_t$ and $y_b$ are the tree-level couplings,
\begin{equation}
  y_t = h_t\cos\b \,, \qquad
  y_b = h_b\sin\b \,,
\end{equation}
with $h_t$ and $h_b$ the top and bottom Yukawa couplings. 
The decay widths at tree level are given by
\begin{equation}
  \G^{\,0}\,(H^\pm \to tb) = \frac{3\kappa}{16\pi\mhp^3}
  \left[ (\mhp^2 - m_t^2 - m_b^2)(y_t^2+y_b^2)
         - 4 m_t m_b y_t y_b \right] \,,
\end{equation}
where $\kappa = \kappa(\mhp^2,m_t^2,m_b^2)$, 
$\kappa(x,y,z)=[(x-y-z)^2-4yz]^{1/2}$.
Since there is no CP violation at tree level, 
$\G^{\,0}(H^+ \to t\bar b)\equiv \G^{\,0}(H^- \to b\bar t)$. 
At next-to-leading order (NLO) we have
\begin{align}
  \G\,(H^\pm \to tb) & =  \frac{3\kappa}{16\pi\mhp^3}
  \left[ (\mhp^2 - m_t^2 - m_b^2)(y_t^2+y_b^2
    + 2y_t\,\Re\,\d Y_t^\pm + 2y_b\,\Re\,\d Y_b^\pm )\right.\nn\\
  & \hspace{24mm}\left. -\, 4 m_t m_b (y_t y_b
    + y_t\,\Re\,\d Y_b^\pm + y_b\,\Re\,\d Y_t^\pm ) \right] \,.
\label{eq:NLO}
\end{align}
The form factors $\d Y_i^\pm$ ($i=t,b$) have, in general, both CP-invariant 
and CP-violating contributions:
\begin{equation}
  \d Y_i^\pm = \d Y_i^{inv} \pm \smaf{1}{2}\,\d Y_i^{CP}\,.
  \label{eq:dYi}
\end{equation}
Both the CP-invariant and the CP-violating contributions have real and
imaginary (absorptive) parts.
CP invariance implies that the form factors of $H^+$ and $H^-$ are equal.
Using Eqs.~\eq{NLO} and \eq{dYi}, we can write the CP-violating asymmetry 
$\d^{CP}$ of Eq.~\eq{defDCP} as \\
\vspace*{-4mm}
\begin{equation}
  \d^{CP} =
  \frac{\Delta\,(y_t\,\Re\,\d Y_t^{CP} + y_b\,\Re\,\d Y_b^{CP})
        - 2 m_t m_b (y_t\,\Re\,\d Y_b^{CP} + y_b\,\Re\,\d Y_t^{CP})}
       {\Delta\,(y_t^2+y_b^2) - 4 m_t m_b\,y_t^{} y_b^{} }\,,
\label{eq:DCP}
\end{equation}

\vspace*{2mm}\noi
where $\Delta = \mhp^2 - m_t^2 - m_b^2 $.

\begin{figure}[ht]
{\setlength{\unitlength}{1mm}
\begin{center}
\begin{picture}(170,40)
\put(10,0){\mbox{\resizebox{!}{35mm}{\includegraphics{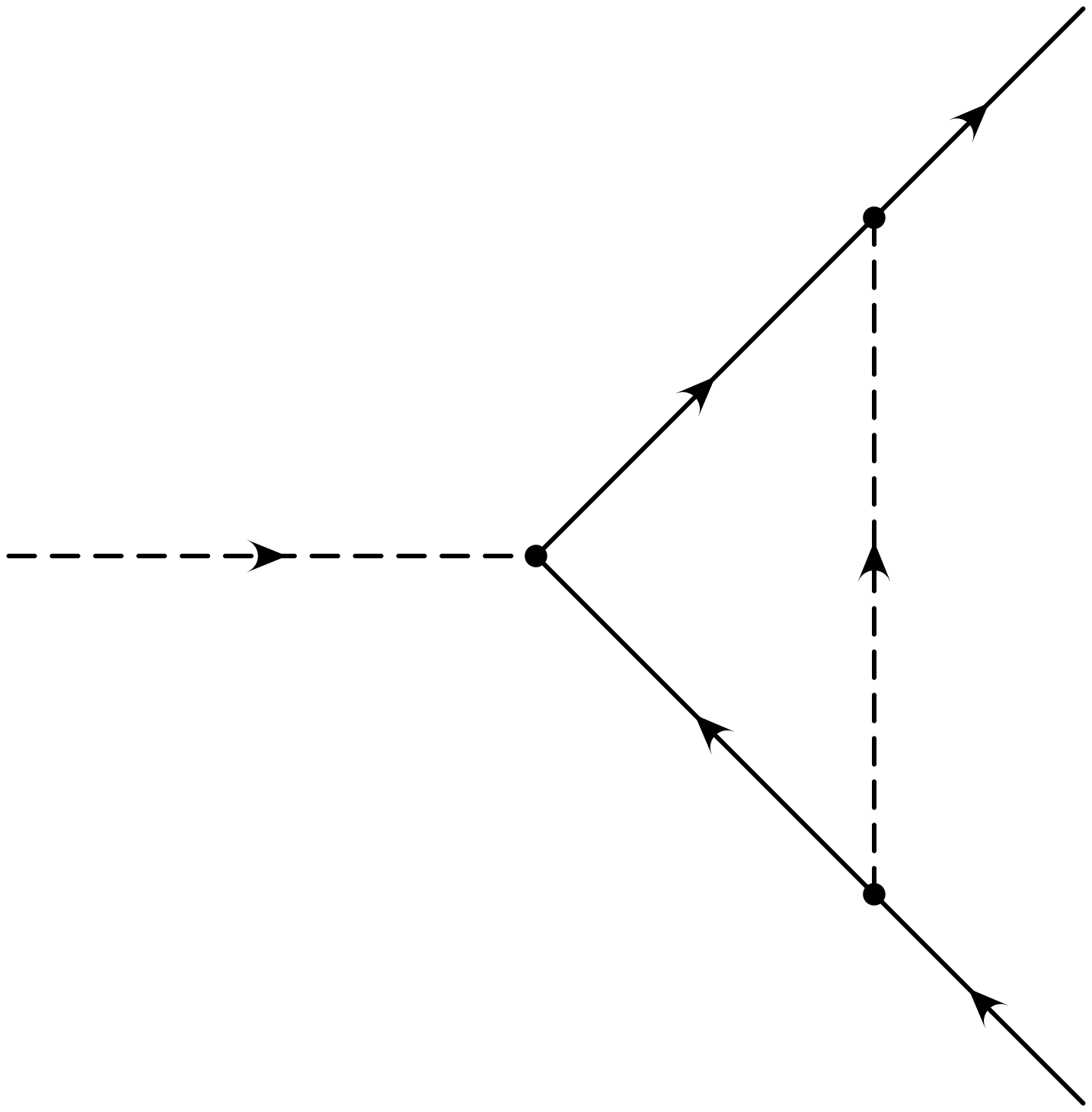}}}}
\put(67,0){\mbox{\resizebox{!}{35mm}{\includegraphics{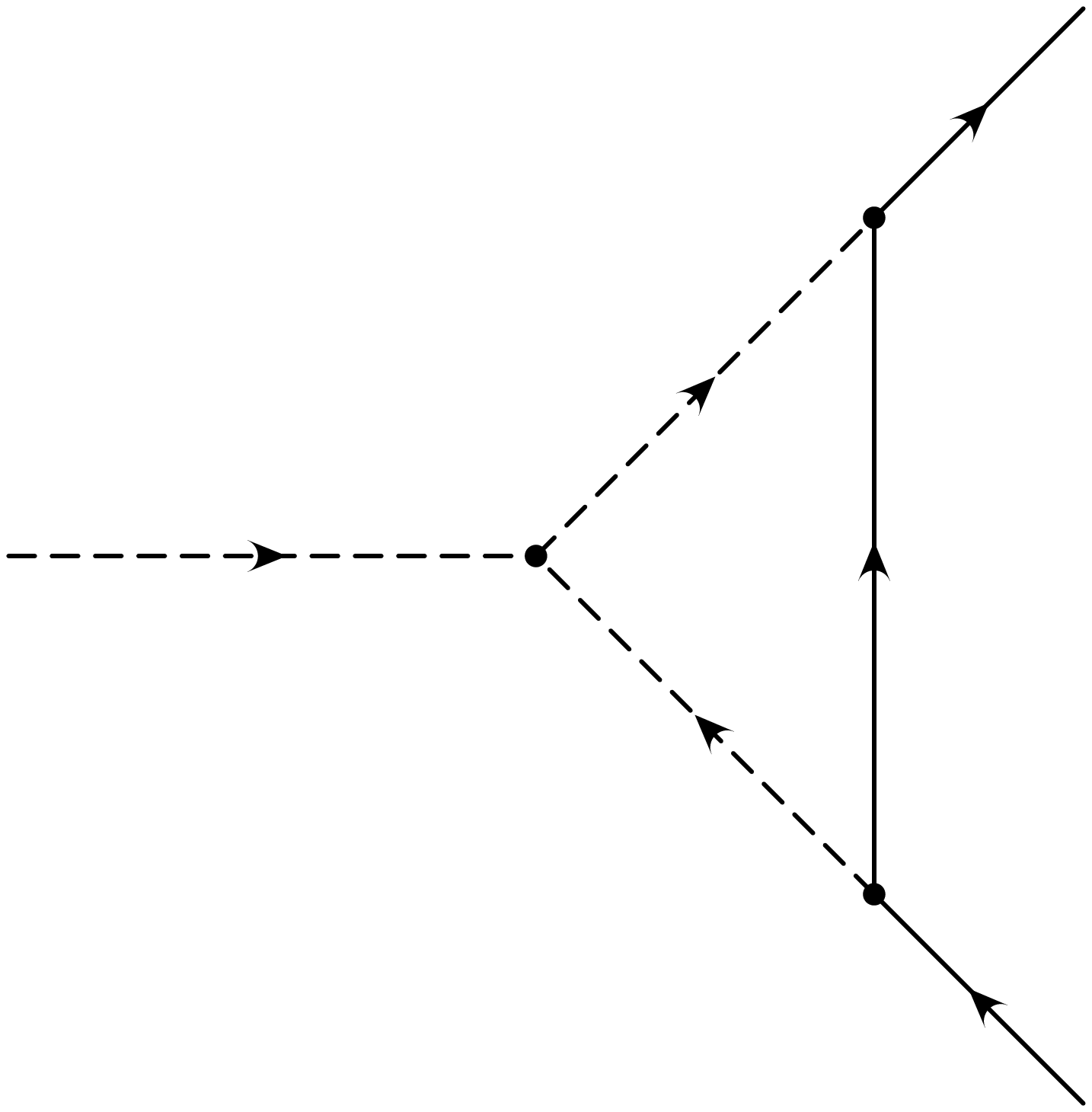}}}}
\put(125,0){\mbox{\resizebox{!}{35mm}{\includegraphics{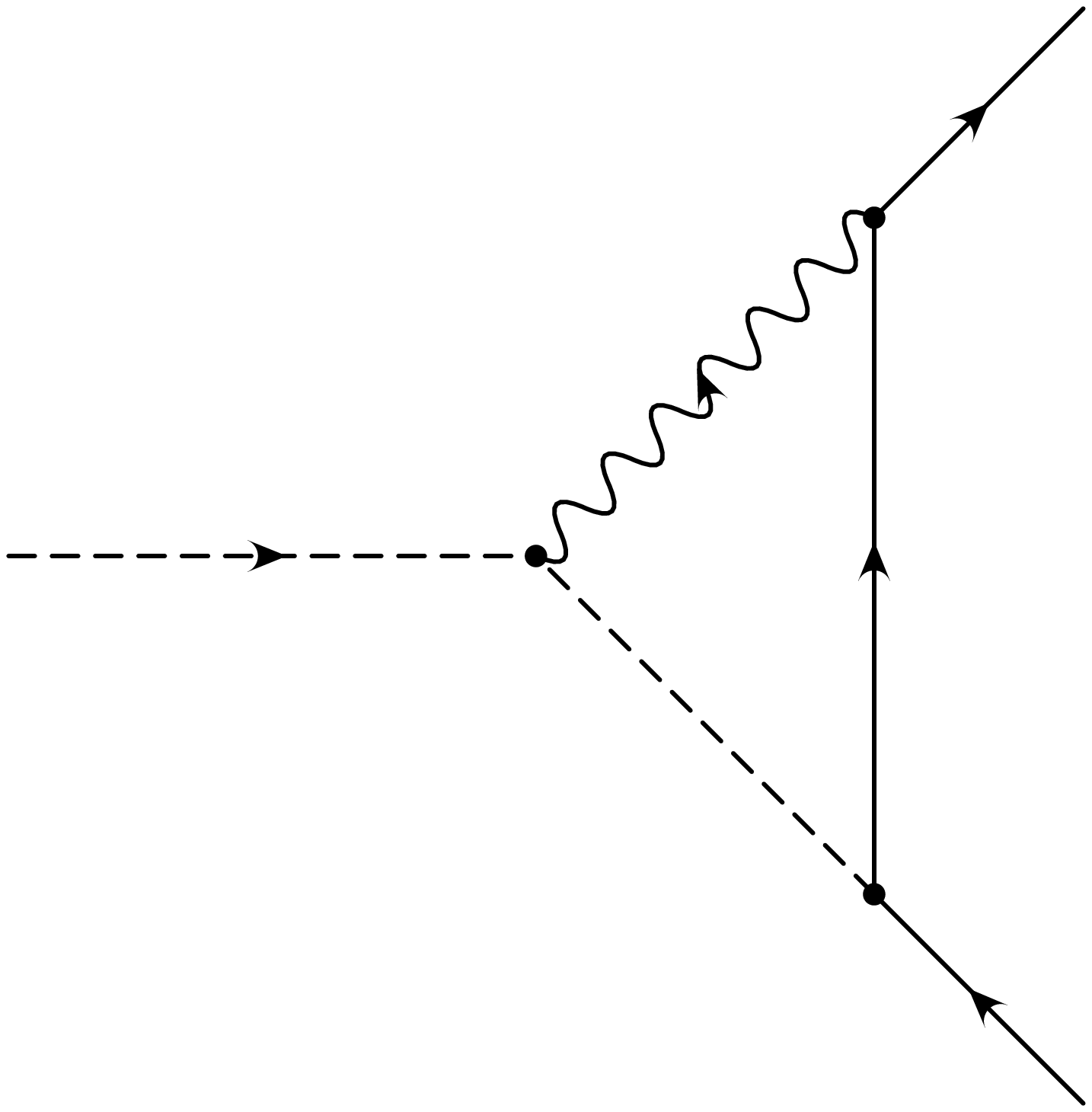}}}}
\put(1,16.5){\mbox{$H^+$}}
\put(58,16.5){\mbox{$H^+$}}
\put(116,16.5){\mbox{$H^+$}}
\put(47,34){\mbox{$t$}}
\put(47,-1){\mbox{$b$}}
\put(104,34){\mbox{$t$}}
\put(104,-1){\mbox{$b$}}
\put(162,34){\mbox{$t$}}
\put(162,-1){\mbox{$b$}}
\put(16,25){\mbox{$\nt_k\;(\chp_j)$}}
\put(16,8){\mbox{$\chm_j\;(\nt_k)$}}
\put(41,16){\mbox{$\st_i^{}\;(\sb_i^{})$}}
\put(83,25){\mbox{$\st_i$}}
\put(83,8){\mbox{$\sb_j$}}
\put(98,16){\mbox{$\nt_k,\,\sg$}}
\put(130,8){\mbox{$H^0_l\;(W)$}}
\put(130,25){\mbox{$W^+\;(H^0_l)$}}
\put(156,16.5){\mbox{$b\;(t)$}}
\put(2,34){\mbox{\bf a)}}
\put(59,34){\mbox{\bf b)}}
\put(117,34){\mbox{\bf c)}}
\end{picture}

\begin{picture}(170,30)
\put(10,0){\mbox{\resizebox{!}{20mm}{\includegraphics{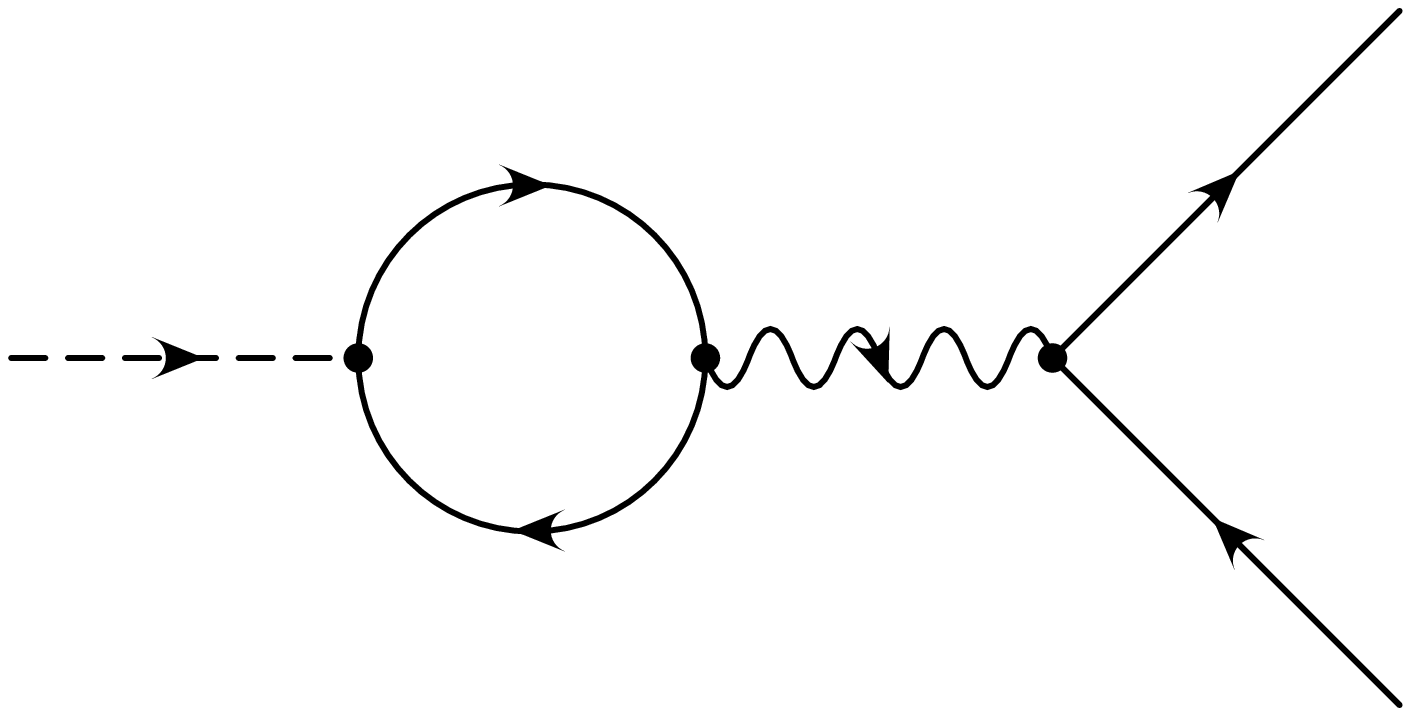}}}}
\put(67,0){\mbox{\resizebox{!}{20mm}{\includegraphics{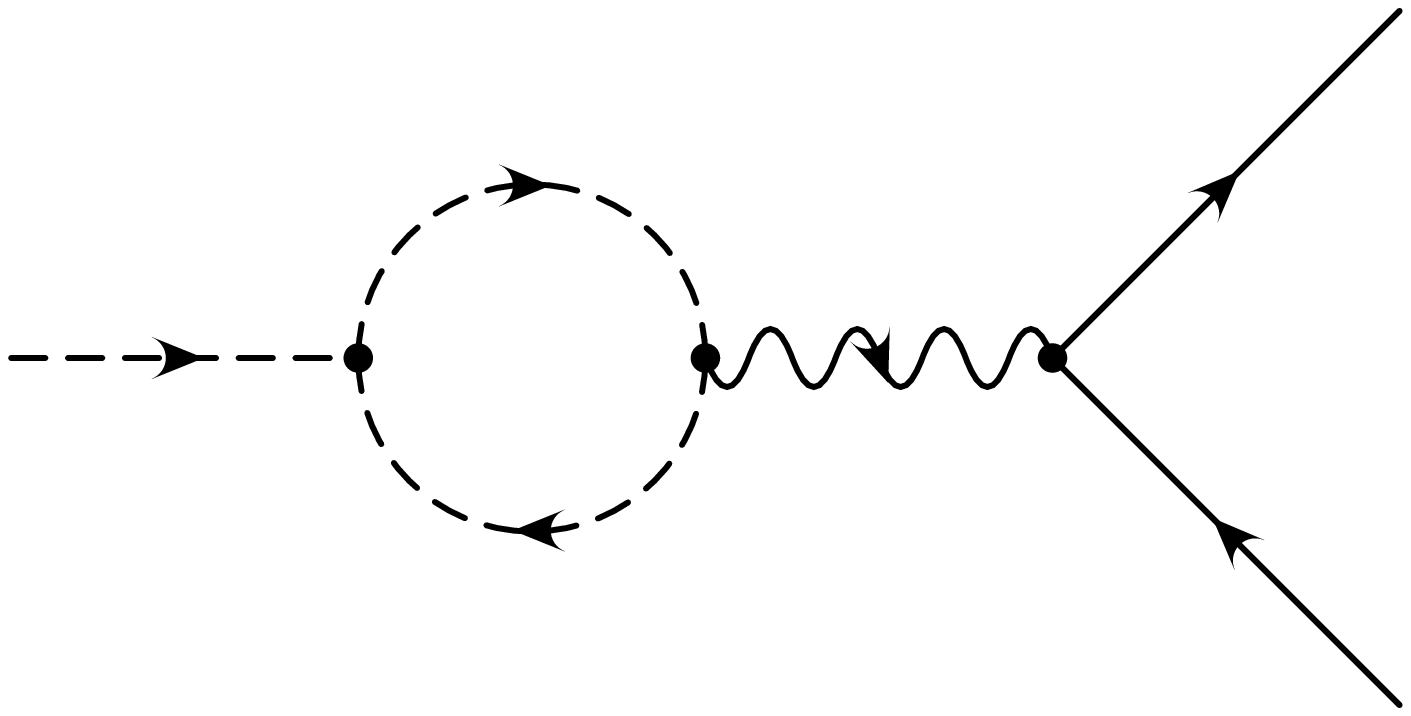}}}}
\put(125,0){\mbox{\resizebox{!}{20mm}{\includegraphics{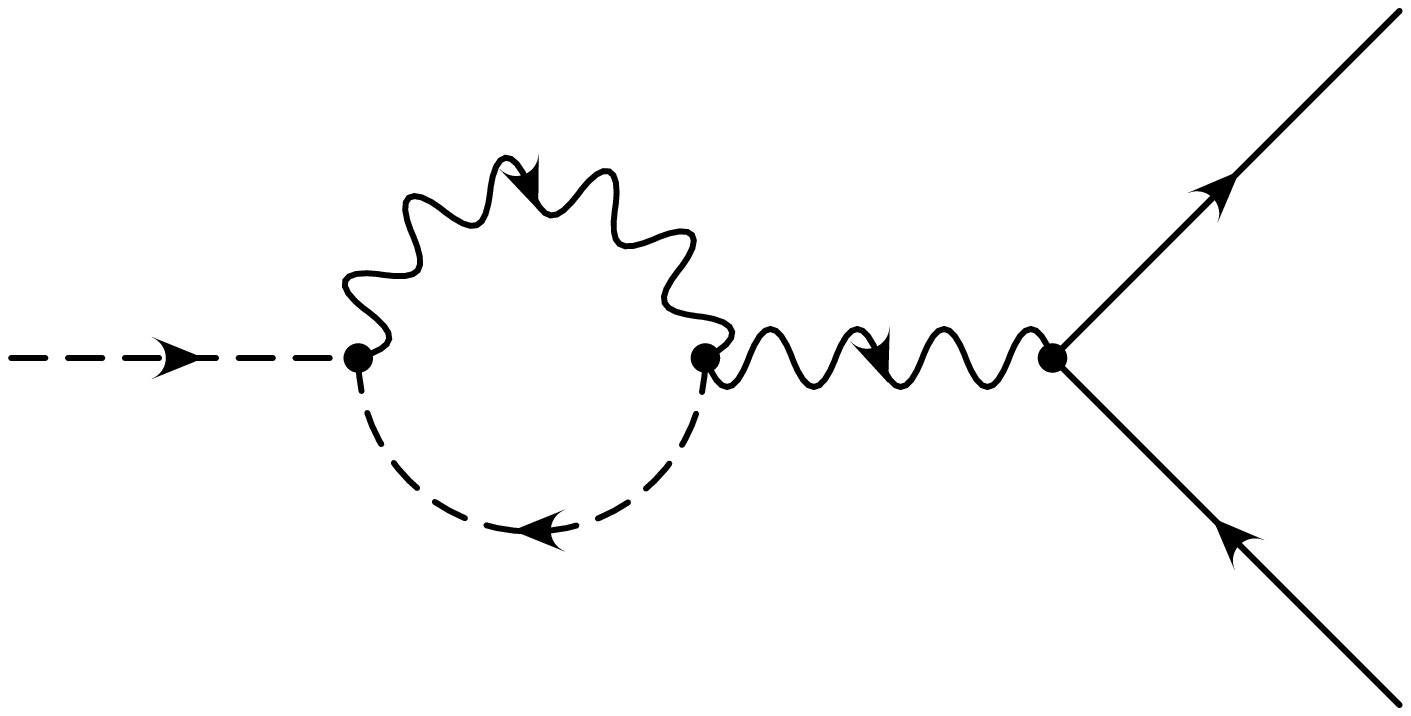}}}}
\put(1,9.5){\mbox{$H^+$}}
\put(58,9.5){\mbox{$H^+$}}
\put(116,9.5){\mbox{$H^+$}}
\put(52,19){\mbox{$t$}}
\put(109,19){\mbox{$t$}}
\put(167,19){\mbox{$t$}}
\put(52,-1){\mbox{$b$}}
\put(109,-1){\mbox{$b$}}
\put(167,-1){\mbox{$b$}}
\put(23.5,18){\mbox{$\nt_k$}}
\put(23,0){\mbox{$\chm_j$}}
\put(33,12.5){\mbox{$W^+$}}
\put(81,17.5){\mbox{$\st_i$}}
\put(81,-0.5){\mbox{$\sb_j$}}
\put(90,12.5){\mbox{$W^+$}}
\put(138,17){\mbox{$W^+$}}
\put(138,-0.5){\mbox{$H^0_l$}}
\put(147.5,12.5){\mbox{$W^+$}}
\put(2,22){\mbox{\bf d)}}
\put(59,22){\mbox{\bf e)}}
\put(117,22){\mbox{\bf f)}}
\end{picture}

\end{center}}
\caption{Sources for CP violation in $H^+\to t\bar b$ decays
at 1-loop level in the MSSM with complex couplings
($i,j=1,2;$ $k=1,...,4$; $l=1,2,3$).
\label{fig:feyngraphs}}
\end{figure}

At one loop, there are six generic diagrams that
may contribute to $\d^{CP}$. These are
(A)~triangle diagrams with
(i)~two fermions and a scalar,
(ii)~two scalars and a fermion, and
(iii)~a scalar, a vector boson and a fermion in the loop, and
(B)~$H^+$--$W^+$ self-energy diagrams with
(i)~two fermions, (ii)~two scalars, and
(iii)~a scalar and a vector boson in the loop.
In the following, we work out the formulae for $\d Y_i^{CP}$
for the specific case of the MSSM with complex phases.
The relevant Feynman diagrams are shown in \fig{feyngraphs}.

\section{CP-violating contributions}

\subsection{Generic diagrams}

According to Eqs.~\eq{MtreeHp} and \eq{Yi} we write the matrix elements
of the 1-loop diagrams of \fig{feyngraphs} as
\begin{equation}
  \M_{H^+} =
  \bar u(p_t)\, \big[ \d Y_b^+\PR + \d Y_t^+\PL\big]\,v(-p_b) \,, 
\end{equation}
and analogously for the $H^-$ decay. 
In fact, we are only interested in the CP-violating parts 
$\Re\,\d Y_i^{CP}$. Since $\Re\,\d Y_i^{CP} = \Re\,(\d Y_i^+ - \d Y_i^-)$,
we just need the structure ${\rm Im}(g_0g_1g_2)\times {\rm Im}(PaV\!e)$
of the form factors $\d Y_b^{+}$ and $\d Y_t^{+}$.
Here $g_0g_1g_2$ stands for the product of the couplings
and $PaV\!e$ for the Passarino--Veltman two- and three-point
functions~\cite{pave} $B_0$ and $C_{0,1,2}$. 
In the following, we give the formulae for the various
contributions to $\Re\,\d Y_{t,b}^{CP}$.
The necessary MSSM mass and mixing matrices, the couplings,
as well as the definition of the two-- and three--point functions
are given in Appendices A, B, and C.

\subsection{Vertex graphs}

\subsubsection{Neutralino--chargino--stop (sbottom) loop}

The graph of Fig.~1a, with a neutralino, a chargino, and a
stop in the loop, leads to
\begin{align}
  &\hspace*{-10mm}\Re\,\d Y^{CP}_b(\nt_k\ch_j\st_i) = \frac{1}{8\pi^2}\,\Big\{
   \left[ m_t m_b\,\Im(F_{jk}^{R}b_{ik}^{\st}l_{ij}^{\st*})
   +\mch{j}\mnt{k}\,\Im(F_{jk}^{R}a_{ik}^{\st}k_{ij}^{\st*})
      + m_t\mch{j}\,\Im(F_{jk}^{R}b_{ik}^{\st}k_{ij}^{\st*}) \right.\nn\\
  &\hspace*{64mm} \left.
     +\;m_b\mnt{k}\,\Im(F_{jk}^{R}a_{ik}^{\st}l_{ij}^{\st*})
      + \mst{i}^2 \,\Im(F_{jk}^{L}a_{ik}^{\st}k_{ij}^{\st*})
   \right]\,\Im (C_0) \nn\\
  &\hspace*{-8mm}
   +m_t \left[ m_t\,\Im(F_{jk}^{L}a_{ik}^{\st}k_{ij}^{\st*})
         + \mnt{k}\,\Im(F_{jk}^{L}b_{ik}^{\st}k_{ij}^{\st*})
             + m_b\,\Im(F_{jk}^{R}b_{ik}^{\st}l_{ij}^{\st*})
         + \mch{j}\,\Im(F_{jk}^{R}b_{ik}^{\st}k_{ij}^{\st*})
   \right]\,\Im (C_1) \nn\\
  &\hspace*{-8mm}
   +m_b \left[ m_t\,\Im(F_{jk}^{R}b_{ik}^{\st}l_{ij}^{\st*}) \,
         + \mnt{k}\,\Im(F_{jk}^{R}a_{ik}^{\st}l_{ij}^{\st*})
             + m_b\,\Im(F_{jk}^{L}a_{ik}^{\st}k_{ij}^{\st*})
         + \mch{j}\,\Im(F_{jk}^{L}a_{ik}^{\st}l_{ij}^{\st*})
   \right]\,\Im (C_2) \nn\\
  & \hspace*{2mm}
   + \Im(F_{jk}^{L}a_{ik}^{\st}k_{ij}^{\st*})\,
     \Im(B_0(\mhp^2,\mnt{k}^2,\mch{j}^2)) \Big\} \,,
  \label{eq:dYntchst}
\end{align}
with $C_X=C_X(m_t^2,\mhp^2,m_b^2,\mst{i}^2,\mnt{k}^2,\mch{j}^2)$, $X=0,1,2$,  
the three-point functions~\cite{pave} in the notation of~\cite{Denner}; 
$\Re\,\d Y^{CP}_t(\nt_k\ch_j\st_i)$ is obtained from Eq.~\eq{dYntchst}
by interchanging $F_{jk}^{L\leftrightarrow R}$,
$a_{ik}^{\st}\leftrightarrow b_{ik}^{\st}$,
$k_{ij}^{\st*}\leftrightarrow l_{ij}^{\st*}$.

The contribution from the neutralino--chargino--sbottom loop has exactly
the same structure. Therefore, $\Re\,\d Y^{CP}_b(\nt_k\ch_j\sb_i)$ is obtained
from Eq.~\eq{dYntchst} by the following substitutions:
for the masses of the loop particles
$\mnt{k} \to \mch{j}$, $\mch{j} \to \mnt{k}$, $\mst{i} \to \msb{i}$;
for the couplings
$a_{ik}^{\st} \to l_{ij}^{\sb}$, $b_{ik}^{\st} \to k_{ij}^{\sb}$,
$k_{ij}^{\st*} \to b_{ik}^{\sb*}$, and $l_{ij}^{\st*} \to a_{ik}^{\sb*}$;
and analogously for $\Re\,\d Y^{CP}_t(\nt_k\ch_j\sb_i)$.

\subsubsection{Stop--sbottom--neutralino loop}

The stop--sbottom--neutralino loop of Fig.~1b gives
\begin{align}
  \Re\,\d Y^{CP}_b(\st_i\,\sb_j\nt_k) = \frac{1}{8\pi^2}\,\big[ \,
  &\mnt{k}\,{\rm Im}(G_{\!4ij}a_{ik}^{\st}b_{jk}^{\sb*})\,{\rm Im}(C_0)\nn\\
  &\hspace{-2mm}
   \,- m_t\,{\rm Im}(G_{\!4ij}b_{ik}^{\st}b_{jk}^{\sb*})\,{\rm Im}(C_1)
   \,- m_b\,{\rm Im}(G_{\!4ij}a_{ik}^{\st}a_{jk}^{\sb*})\,{\rm Im}(C_2)
   \,\big] \,, \\
  \Re\,\d Y^{CP}_t(\st_i\,\sb_j\nt_k) = \frac{1}{8\pi^2}\,\big[ \,
  &\mnt{k}\,{\rm Im}(G_{\!4ij}b_{ik}^{\st}a_{jk}^{\sb*})\,{\rm Im}(C_0)\nn\\
  &\hspace{-2mm}
   \,- m_t\,{\rm Im}(G_{\!4ij}a_{ik}^{\st}a_{jk}^{\sb*})\,{\rm Im}(C_1)
   \,- m_b\,{\rm Im}(G_{\!4ij}b_{ik}^{\st}b_{jk}^{\sb*})\,{\rm Im}(C_2)
   \,\big] \,,
\end{align}
with $C_X=C_X(m_t^2,\mhp^2,m_b^2,\mnt{k}^2,\mst{i}^2,\msb{j}^2)$.

\subsubsection{Stop--sbottom--gluino loop}

The contribution from the diagram with a stop, a sbottom, and
a gluino in Fig.~1b is
\begin{align}
  \Re\,\d Y^{CP}_b(\st_i\,\sb_j\sg) = -\frac{4}{3}\frac{\a_s}{\pi}\,
    & \big[ \,
      \msg\,\Im(G_{4ij}\Rsts_{1i}\Rsb_{2j}\,e^{i\phi_3})\,\Im(C_0) \nn\\
    & \hspace{-5mm}
      + m_t\,\Im(G_{4ij}\Rsts_{2i}\Rsb_{2j})\,\Im(C_1)
      + m_b\,\Im(G_{4ij}\Rsts_{1i}\Rsb_{1j})\,\Im(C_2) \,\big]\,,\\
  \Re\,\d Y^{CP}_t(\st_i\,\sb_j\sg) = -\frac{4}{3}\frac{\a_s}{\pi}\,
    & \big[ \,
      \msg\,\Im(G_{4ij}\Rsts_{2i}\Rsb_{1j}\,e^{-i\phi_3})\,\Im(C_0) \nn\\
    & \hspace{-5mm}
      + m_t\,\Im(G_{4ij}\Rsts_{1i}\Rsb_{1j})\,\Im(C_1)
      + m_b\,\Im(G_{4ij}\Rsts_{2i}\Rsb_{2j})\,\Im(C_2) \,\big]\,,
\end{align}
with $C_X=C_X(m_t^2,\mhp^2,m_b^2,\msg^2,\mst{i}^2,\msb{j}^2)$
and $\a_s=g_s^2/(4\pi)$.

\subsubsection{W boson--neutral Higgs--bottom (top) loop}

There are two contributions, one with a bottom and one
with a top quark in the loop (with $H^0_l$ and $W$ interchanged),
see Fig.~1c. We use the $\xi=1$ gauge. The $WH_l\,b$ loop gives:
\begin{align}
  \Re\,\d Y^{CP}_b(W H_l b) = \,
    & -\,\frac{\sqrt{2}\,g^2}{32\pi^2}\, \Big\{ \Im(X_b^R)\,
      \big[ (3 m_b^2 - 2 m_{H_l}^2)\,\Im(C_0)
            + m_t^2\,\Im(C_1) + 2 m_b^2\,\Im(C_2) \nn\\
    & \hspace{18mm}
            +\,\Im\big(B_0(m_{H^+}^2, m_W^2, m_{H_l}^2)\big)
            - 2\,\Im\big(B_0(m_t^2, m_b^2, m_W^2)\big)\big] \nn\\
    & \hspace{18mm}
            +\,m_b^2\,\Im(X_b^L)\,\Im( 2 C_0 + C_2) \Big\} \,,
    \label{eq:dY1CP5} \\
  \Re\,\d Y^{CP}_t(W H_l b) = \,
    & -\,\frac{\sqrt{2}\,g^2}{32\pi^2}\,\,m_t\,m_b\,\left[
      \Im(X_b^R)\,\Im(2C_1 + C_2) + \Im(X_b^L)\,\Im(C_1 - C_0) \right] \,,
    \label{eq:dY2CP5}
\end{align}
where $X_b^R = g_{H_lH^+W^-}^{}\,s_l^{b,R}$,
$X_b^L = g_{H_lH^+W^-}^{}\,s_l^{b,L}$, and
$C_X=C_X(m_t^2,\mhp^2,m_b^2,m_b^2,m_W^2, m_{H_l}^2)$.
Analogously, the $H_l W t$ loop gives
\begin{align}
  \Re\,\d Y^{CP}_b(H_l W t) = \,
    & \frac{\sqrt{2}\,g^2}{32\pi^2}\,\,m_t\,m_b\,\left[
      \Im(X_t^L)\,\Im(2C_1 + C_2) + \Im(X_t^R)\,\Im(C_1 - C_0) \right]\,,
  \label{eq:dY1CP6} \\
  \Re\,\d Y^{CP}_t(H_l W t) = \,
    & \frac{\sqrt{2}\,g^2}{32\pi^2}\, \Big\{ \Im(X_t^L)\,
      \big[ (3 m_t^2 - 2 m_{H_l}^2)\,\Im(C_0)
            + m_b^2\,\Im(C_1) + 2 m_t^2\,\Im(C_2) \nn\\
    & \hspace{15mm}
            +\,\Im\big(B_0(m_{H^+}^2, m_W^2, m_{H_l}^2)\big)
            - 2\,\Im\big(B_0(m_b^2, m_t^2, m_W^2)\big)\big] \nn\\
    & \hspace{15mm}
            +\,m_t^2\,\Im(X_t^R)\,\Im( 2 C_0 + C_2) \Big\}\,,
  \label{eq:dY2CP6}
\end{align}
with $X_t^R = g_{H_lH^+W^-}^{}\,s_l^{t,R}$,
$X_t^L = g_{H_lH^+W^-}^{}\,s_l^{t,L}$, and
$C_X=C_X(m_t^2,\mhp^2,m_b^2,m_t^2,m_{H_l}^2,m_W^2)$.

\subsubsection{Ghost--neutral Higgs--bottom (top) loop}

Since the above graphs with a $W$ boson are calculated in the
$\xi=1$ gauge, we also have to include the corresponding graphs
with $W^\pm\to G^\pm$. These lead to
\begin{align}
  \Re\,\d Y^{CP}_b(G H_l b) = \,
  & -\frac{1}{8\pi^2}\, \left[ \,
    m_b h_b\cos\b\,\Im\,(\hat X_b^R)\,\Im\,(C_0) +
    m_t h_t\sin\b\,\Im\,(\hat X_b^R)\,\Im\,(C_1) \right.\nn\\
  & \left.\hspace{5cm}
    -\,m_b h_b\cos\b\,\Im\,(\hat X_b^L)\,\Im\,(C_2)\,\right] \,,\\
  \Re\,\d Y^{CP}_t(G H_l b) = \,
  & \frac{1}{8\pi^2}\, \left[ \,
    m_b h_t\sin\b\,\Im\,(\hat X_b^L)\,\Im\,(C_0) +
    m_t h_b\cos\b\,\Im\,(\hat X_b^L)\,\Im\,(C_1) \right.\nn\\
  & \left.\hspace{5cm}
    -\,m_b h_t\sin\b\,\Im\,(\hat X_b^R)\,\Im\,(C_2)\,\right] \,,
\end{align}
and
\begin{align}
  \Re\,\d Y^{CP}_b(H_l G\,t) = \,
  & -\frac{1}{8\pi^2}\, \left[ \,
    m_t h_b\cos\b\,\Im\,(\hat X_t^R)\,\Im\,(C_0) -
    m_t h_b\cos\b\,\Im\,(\hat X_t^L)\,\Im\,(C_1) \right.\nn\\
  & \left.\hspace{5cm}
    +\,m_b h_t\sin\b\,\Im\,(\hat X_t^R)\,\Im\,(C_2)\,\right] \,,\\
  \Re\,\d Y^{CP}_t(H_l G\,t) = \,
  & \frac{1}{8\pi^2}\, \left[ \,
    m_t h_t\sin\b\,\Im\,(\hat X_t^L)\,\Im\,(C_0) -
    m_t h_t\sin\b\,\Im\,(\hat X_t^R)\,\Im\,(C_1) \right.\nn\\
  & \left.\hspace{5cm}
    +\,m_b h_b\cos\b\,\Im\,(\hat X_t^L)\,\Im\,(C_2)\,\right] \,.
\end{align}

\noi
Here, $\hat X_q^{R,L} = g_{H_lH^+G^-}^{}\,s_l^{q\,R,L}$ for $q=b,t$.
The $C$ functions are $C_X=C_X(m_t^2,\mhp^2,m_b^2,m_b^2,m_W^2,m_{H_l}^2)$
for $q=b$ and $C_X=C_X(m_t^2,\mhp^2,m_b^2,m_t^2,m_{H_l}^2,m_W^2)$
for $q=t$.

\subsection{Self-energy graphs}

\subsubsection{Neutralino--chargino loop}

The self-energy graph with a neutralino and a chargino of Fig.~1d gives
\begin{align}
  \Re\,\d Y^{CP}_{b\,(t)}\,(\nt_k\ch_j - W)  =\,
  & \pm\,\frac{1}{8 \pi^2}\,
    \frac{g^2\, m_{b\,(t)}}{\sqrt{2}\,m_{H^+}^2\, m_W^2}\,
    \Im\left(B_0(\mhp^2, \mnt{k}^2, \mch{j}^2)\right) \, \times
    \nonumber \\
  & \hspace*{-24mm} \left[
    \Im\left(c_{II}\right)\mch{j} (\mhp^2 + \mnt{k}^2 - \mch{j}^2) -
    \Im\left(c_{IJ}\right)\mnt{k} (\mhp^2 - \mnt{k}^2 + \mch{j}^2)
    \right]
\label{eq:dY12ntch-W}
\end{align}
with
$c_{II} = F_{jk}^R O_{kj}^R + F_{jk}^L O_{kj}^L\,$, and
$c_{IJ} = F_{jk}^R O_{kj}^L + F_{jk}^L O_{kj}^R$.
The overall plus sign is for $\d Y^{CP}_{b}$, and
the overall minus sign for $\d Y^{CP}_{t}$.

\subsubsection{Stop--sbottom loop}

The graph of Fig.~1e leads to
\begin{align}
  \Re\,\d Y^{CP}_{b\,(t)}\,(\st_i\,\sb_j - W) =\;
  & \mp\,\frac{3g^2}{16\pi^2}\,
    \frac{m_{b\,(t)}}{\mhp^2 m_W^2}\,(\mst{i}^2-\msb{j}^2)\,
    \times \nn\\
  & \hspace{26mm}
    \Im\,(G_{\!4ij}\Rst_{1i}\Rsbs_{1j})\;
    \Im \left(B_0(\mhp^2,\msb{j}^2,\mst{i}^2)\right)\,.
\end{align}

\subsubsection{\boldmath $W^\pm$--$H^0_l$ and $G^\pm$--$H^0_l$ loops}

The self-energy graph with $W^+$ and $H^0_l$ is shown in Fig.~1f.
Since we use $\xi=1$ gauge for the $W$ in the loop,
we have to add the corresponding graph with a ghost,
i.e. $W^\pm\to G^\pm$ in the loop.
(The second $W$ propagator can be calculated in the unitary gauge.
Hence, no ghost is necessary in this case.)
The two contributions together give:
\begin{align}
  \Re\,\d Y^{CP}_{b\,(t)}\,(W H^0_l - W) =\;
  & \mp\,\frac{1}{32\pi^2}\,
    \frac{g^3\,m_{b\,(t)}}{\sqrt{2}\,\mhp^2 m_W^{}}\,
    (2m_W^2 - 2m_{H_l}^2 - 3\,\mhp^2)\, \times \nn\\
  & \hspace{1cm}
    O_{3l}\,(\cos\b\,O_{1l} + \sin\b\,O_{2l})~
    \Im\left(B_0(\mhp^2,m_{H_l}^2,m_W^2)\right)\,.
\end{align}

\section{Numerical results}

Let us now turn to the numerical analysis.
In order not to vary too many parameters, we fix part
of the parameter space at the electroweak scale by the choice 
\footnote{In the original version of the paper \cite{published}  
there were two mistakes in the analytic expressions~\cite{erratum}, 
which also affected the numerical analysis: 
for $\tan\b=10$ (40) the results for $\d^{CP}$ 
in \cite{published} are reduced to about 30\% (50\%).  
However, one can easily find scenarios where $\d^{CP}$ goes up 
to 10--15\%. In particular, for the parameters of Eq.~\eq{parset}, 
the resulting plots are very similar to the original ones.}
\begin{align}
   & M_2=200~{\rm GeV},~ \mu=-350~{\rm GeV},~
     M_{\ti U} : M_{\ti Q} : M_{\ti D} = 0.85 : 1 : 1.05, \nn\\
   & A_t = A_b = -500~{\rm GeV}.
\label{eq:parset}
\end{align}
Moreover, we assume GUT relations for the gaugino mass parameters
$M_1$, $M_2$, $M_3$. In this case, the phases of the gaugino
sector can be rotated away. 
Since $\phi_\mu$, the phase of $\mu$, is highly constrained by
the EDMs of electron and neutron, we take $\phi_\mu = 0$.
The phases relevant to our study are thus
$\phi_t$ and $\phi_b$, the phases of $A_t$ and $A_b$.

\begin{figure}[p]
\setlength{\unitlength}{1mm}
\begin{center}
\begin{picture}(70,62)
\put(0,-2){\mbox{\epsfig{figure=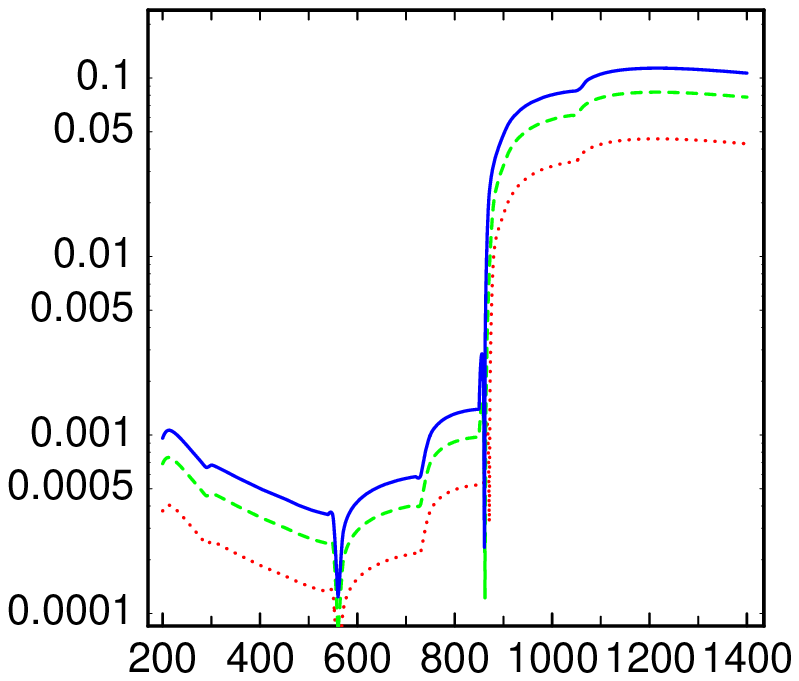,height=70mm}}}
\put(30,-2){$m_{H^+}$~[GeV]}
\put(-6,30){\rotatebox{90}{$|\d^{CP}|$}}
\end{picture}
\caption{Absolute value of $\d^{CP}$ as a function of $\mhp$,
  for $M_{\ti Q}=490$~GeV and $\tan\b=10$.
  The solid, dashed, and dotted lines are for
  $\phi_t=\frac{\pi}{2},~\frac{\pi}{4},$ and $\frac{\pi}{8}$,
  respectively, 
  $\phi_b=0$, the other parameters are fixed by Eq.~\eq{parset}.
\label{fig:S1mH}}
\end{center}
\end{figure}

The choice in Eq.~\eq{parset} together with $M_{\ti Q}=490$~GeV
and $\tan\b=10$ gives a mass spectrum quite similar to the Snowmass
point SPS1a~\cite{SPS}.
\Fig{S1mH} shows $\d^{CP}$ for this case as a function of $\mhp$
in the range $\mhp = 200$\,--\,1400~GeV
and various values of $\phi_t$.
The sparticle masses are given explicitly in Table~1.

\begin{table}
\begin{center}
\begin{tabular}{r}
\begin{tabular}{|c||cccc|cc|}
\hline
  $\tan\b$
  & $\mnt{1}$ & $\mnt{2}$ & $\mnt{3}$ & $\mnt{4}$ & $\mch{1}$ & $\mch{2}$ \\
\hline
  10 & 99 & 191 & 359 & 369 & 191 & 372 \\
  40 & 98 & 188 & 358 & 372 & 188 & 374 \\
\hline
\end{tabular} \\[12mm]

\begin{tabular}{|cc||cc|cc|}
\hline
  $M_{\ti Q}$ & $\tan\b$
   & \hspace*{6mm}$\mst{1}$\hspace*{6mm} & \hspace*{6mm}$\mst{2}$\hspace*{6.3mm}
   & \hspace*{0.5mm}$\msb{1}$\hspace*{0.5mm} & \hspace*{0.5mm}$\msb{2}$\hspace*{1mm}\\
\hline
  490 & 10 & 384 (377) & 568 (573) & 486 & 522 \\
      & 40 & 379 (377) & 571 (573) & 435 & 566 \\
\hline
  350 & 10 & 226 (213) & 465 (471) & 340 & 382 \\
      & 40 & 216 (212) & 470 (472) & 257 & 443 \\
\hline
\end{tabular}
\end{tabular}
\end{center}
\caption{Sparticle masses (in GeV) for parameter sets used
  in the numerical analysis for $\phi_t=0~(\frac{\pi}{2})$.}
\end{table}

For $\mhp < \mst{1}+\msb{1}$, $\d^{CP}$ is very small,
${\cal O}(10^{-3})$ or smaller.
The contributions to $\d^{CP}$ come from the diagrams of
Figs.~1a, 1c, and 1f; the diagram of \fig{feyngraphs}d only contributes 
if there is a non-zero phase in the chargino/neutralino sector. 
In the region $\mhp = 200-800$ GeV, one can distinguish the thresholds of
$\nt_1\ch_1$ at $\mhp\simeq 290$~GeV, $\nt_1\ch_2$ at $\mhp\simeq 470$~GeV,
$\nt_3\ch_1$ at $\mhp\simeq 550$~GeV, $\nt_2\ch_2$ and $\nt_4\ch_1$
at $\mhp\simeq 560$~GeV, and $\nt_{3,4}\ch_2$ at $\mhp\simeq 730$--740~GeV.
Below the $\nt_1\ch_1$ threshold, $\d^{CP}$ originates only from the graphs
with $W$ and a neutral Higgs boson of Figs.~1c and 1f. Here note
that the graph with $W H^0_l\,b$ of Fig.~1c always contributes, since
$m_t>m_W+m_b$. 

However, once the $H^+\to\st\bar{\sb}$ channel is open,
$\d^{CP}$ can go up to several per cent.
The thresholds of $H^+\to\st_1\bar{\sb}_1$ at $\mhp\sim 860$~GeV,
and of $H^+\to\st_2\bar{\sb}_2$ at $\mhp\sim 1100$~GeV
are clearly visible in \fig{S1mH}. 
For $\mhp=1$~TeV, we obtain $\d^{CP}\simeq -3\%$, $-6\%$, and $-8\%$ 
for $\phi_t=\frac{\pi}{8}$, $\frac{\pi}{4}$, and $\frac{\pi}{2}$, 
respectively.
The dominant contribution comes from the stop--sbottom--gluino
loop of \fig{feyngraphs}b.
Also the stop--sbottom--neutralino loop of Fig.~1b and the
stop--sbottom self-energy of Fig.~1e can give a relevant
contribution and should thus be taken into account. The contribution
of the graphs with $\ch\nt$ or $H^0W$ (\fig{feyngraphs}a,c,f) exchange
can, however, be neglected in this case.

%
%

\begin{figure}[p]
\setlength{\unitlength}{1mm}
\begin{center}
\begin{picture}(60,62)
\put(-2,-2){\mbox{\epsfig{figure=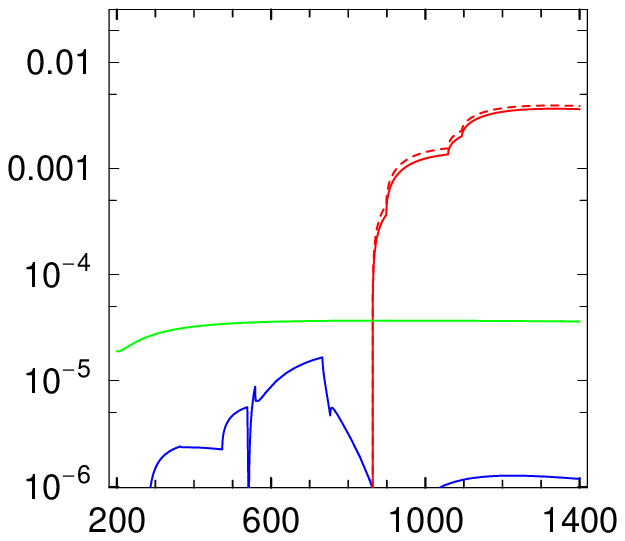,height=70mm}}}
\put(43,49){$\st\sb$}
\put(52,30){\footnotesize $H^0W$}
\put(17,19.5){\footnotesize $\ch\nt$}
\put(30,-1){$m_{H^+}$~[GeV]}
\put(-8,28){\rotatebox{90}{$|\Re\,\d Y_b^{CP}|$}}
\put(-7,58){\bf a)}
\end{picture}
\hspace{16mm}
\begin{picture}(60,62)
\put(0,-2){\mbox{\epsfig{figure=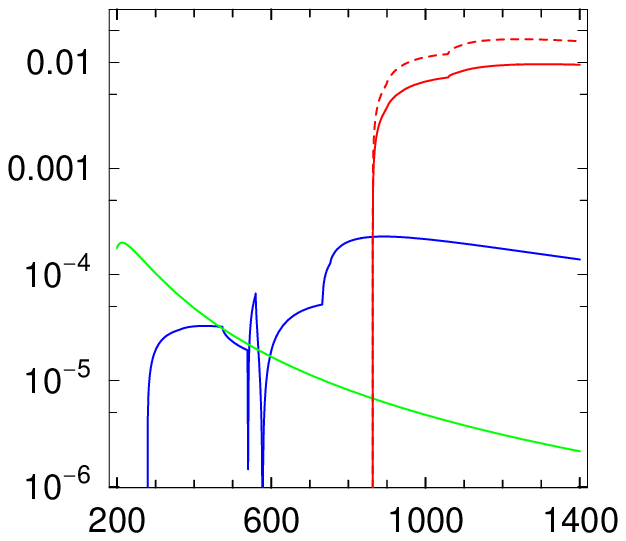,height=70mm}}}
\put(36,52){$\st\sb$}
\put(52,18){\footnotesize $H^0W$}
\put(52,39){\footnotesize $\ch\nt$}
\put(30,-1){$m_{H^+}$~[GeV]}
\put(-6,28){\rotatebox{90}{$|\Re\,\d Y_t^{CP}|$}}
\put(-6,58){\bf b)}
\end{picture}
\caption{Absolute values of $\Re\,\d Y_b^{CP}$ and $\Re\,\d Y_t^{CP}$
  as functions of $\mhp$, for $\phi_t=\pi/2$ and the 
  other parameters as in \fig{S1mH}.
  The blue lines show the contribution from the $\ch\nt$ exchanges
  of Fig.~1a; the red lines are those from the $\st\sb$
  exchanges of Figs.~1b and 1e;
  the green lines are those from the diagrams with
  $H^0$ and $W$ of Figs.~1c and 1f.
  The red dashed lines show $\Re\,\d Y_i^{CP}$ due to the 
  $\st\sb\sg$ loop only. \label{fig:dY12}}
\end{center}
\end{figure}

The relative importance of the various contributions is illustrated
in \fig{dY12}, where we plot the form factors $\Re\,\d Y_b^{CP}$ and
$\Re\,\d Y_t^{CP}$ as functions of $\mhp$, for $\phi_t=\pi/2$ and the 
other parameters as in \fig{S1mH}.
To calculate the contributions with neutral Higgs bosons, we have used 
\cite{CEPW00,cph}. This is sufficient for our purpose, since we are 
mainly interested in large CP-violating effects that occur for 
$\mhp>\mst{1}+\msb{1}$ because of $\phi_{t,b}$. 
However, once precision measurements of $H^\pm$ decays become feasible, 
a more complete calculation of the $H^0_l$ masses and couplings 
\cite{Carena:2001fw} might be used.

\begin{figure}[p]
\setlength{\unitlength}{1mm}
\begin{center}
\begin{picture}(60,62)
\put(0,0){\mbox{\epsfig{figure=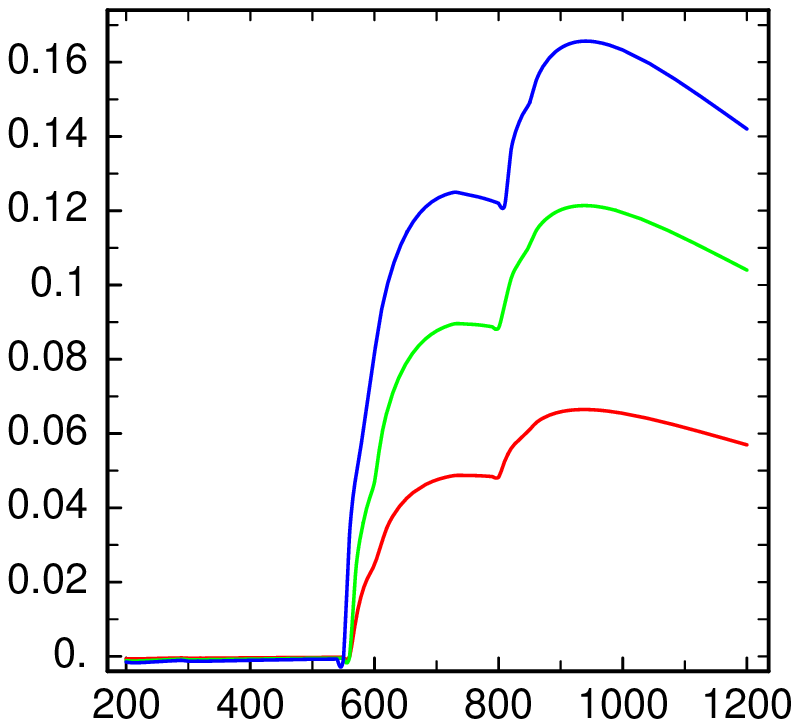,height=61mm}}}
\put(24,49){\footnotesize $\phi_t=\pi/2$}
\put(49,43.5){\footnotesize $\pi/4$}
\put(49,28.5){\footnotesize $\pi/8$}
\put(24,-2){$m_{H^+}$~[GeV]}
\put(-6,28){\rotatebox{90}{$-\d^{CP}$}}
\put(-6,56){\bf a)}
\end{picture}
\hspace{16mm}
\begin{picture}(60,62)
\put(0,0){\mbox{\epsfig{figure=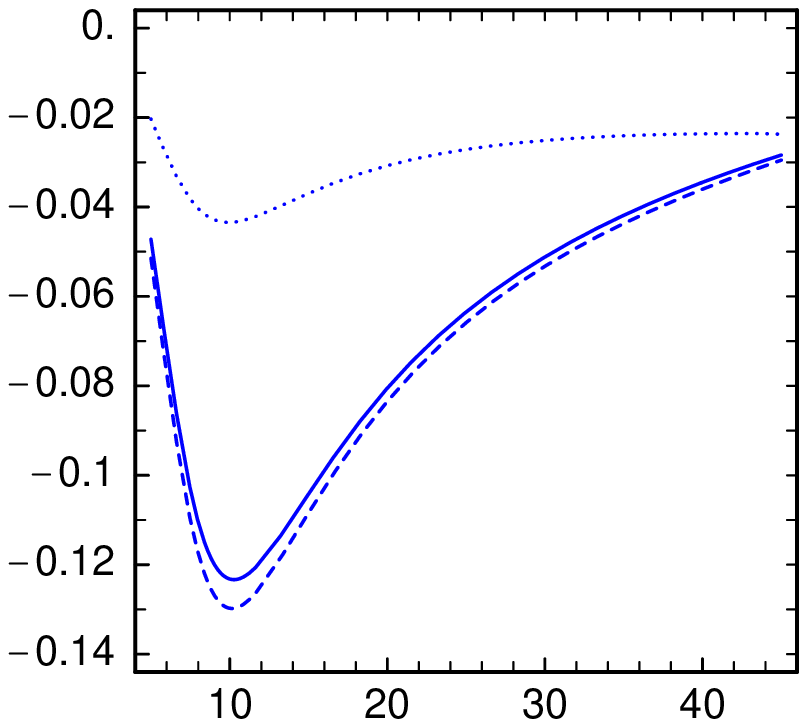,height=63mm}}}
\put(34,51.5){\tiny $\mu=+350$~GeV, $\phi_b=0$}
\put(25,36){\footnotesize $\phi_b=0$}
\put(25.5,16){\footnotesize $\phi_b=\frac{\pi}{2}$}
\put(34,-2){$\tan\b$}
\put(-5,30){\rotatebox{90}{$\d^{CP}$}}
\put(-5,56){\bf b)}
\end{picture}
\caption{$\d^{CP}$ for $M_{\ti Q}=350$~GeV;
  in {\bf (a)} as a function of $\mhp$, for $\tan\b=10$, $\phi_b=0$, and
  in {\bf (b)} as a function of $\tan\b$, for $\mhp=700$~GeV 
  and $\phi_t=\frac{\pi}{2}$.
  The other parameters are fixed by Eq.~\eq{parset}.
  \label{fig:S2mHtb}}
\end{center}
\end{figure}

We next lower the stop/sbottom mass scale to $M_{\ti Q}=350$~GeV.
The resulting masses are given in Table~1.
Figures~\ref{fig:S2mHtb}a and \ref{fig:S2mHtb}b show $\d^{CP}$
for this case as functions of $\mhp$ and $\tan\b$, respectively.
The threshold behaviour of \fig{S2mHtb}a is very similar to
that of \fig{S1mH}a. The threshold of $H^+\to\st_1\sb_1$ is shifted 
to $\mhp\simeq 550$~GeV, and $\d^{CP}$ reaches larger values 
for lighter squarks. Even for a small phase $\phi_t$, $\d^{CP}$ 
can be a few per cent. 
\Fig{S2mHtb}b shows the $\tan\b$ dependence of $\d^{CP}$
for $\mhp=700$~GeV and the cases
$\phi_t=\pi/2$, $\phi_b=0$ (full line) and $\phi_t=\phi_b=\pi/2$
(dashed line). For completeness, we also show as a dotted line
the case of $\mu = +350$~GeV (with $\phi_t=\pi/2$, $\phi_b=0$).
It turns out that the asymmetry has a maximum around 
$\tan\b\simeq 10$ and decreases for larger values of $\tan\b$. 
In particular, we have $\d^{CP}\sim -12\%$ ($-3.5\%$) for 
$\tan\b=10$ (40), $\phi_t=\pi/2$, $\phi_b=0$, and $\mu=-350$~GeV. 
An additional phase of $A_b$ can enhance or reduce the asymmetry.  
For $\mu<0$, however, its effects in the triangle and 
self-energy graphs of Fig.~1b and 1e compensate each other 
so that the overall dependence on $\phi_b$ is small. 
The dependence is larger for $\mu>0$.
 
Here we also note that the branching ratio of $H^+\to t\bar b$
increases with $\tan\b$. In the case of vanishing phases
we have BR$(H^+\to t\bar b)\simeq 17\%$ (15\%) at $\tan\b=10$
and 85\% (75\%) at $\tan\b=40$ for $M_{\ti Q} = 350$ (490)~GeV 
and $\mhp=700$ (1000)~GeV.

\begin{figure}[p]
\setlength{\unitlength}{1mm}
\begin{center}
\begin{picture}(60,62)
\put(0,0){\mbox{\epsfig{figure=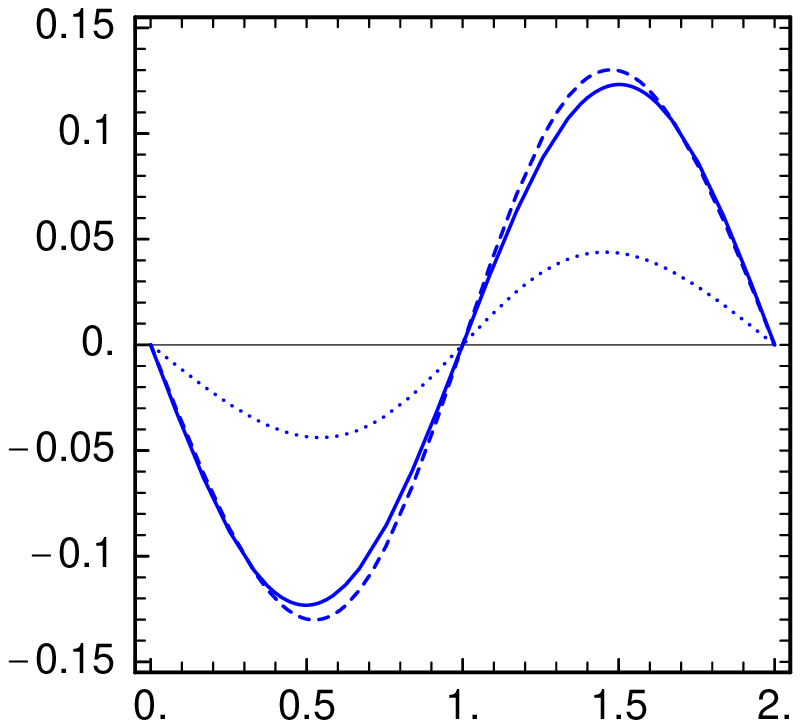,height=63mm}}}
\put(40,13){\footnotesize\mbf $\tan\b=10$}
\put(32,-2){$\phi_t~[\pi]$}
\put(-5,31){\rotatebox{90}{$\d^{CP}$}}
\put(-5,56){\bf a)}
\end{picture}
\hspace{16mm}
\begin{picture}(60,62)
\put(0,0){\mbox{\epsfig{figure=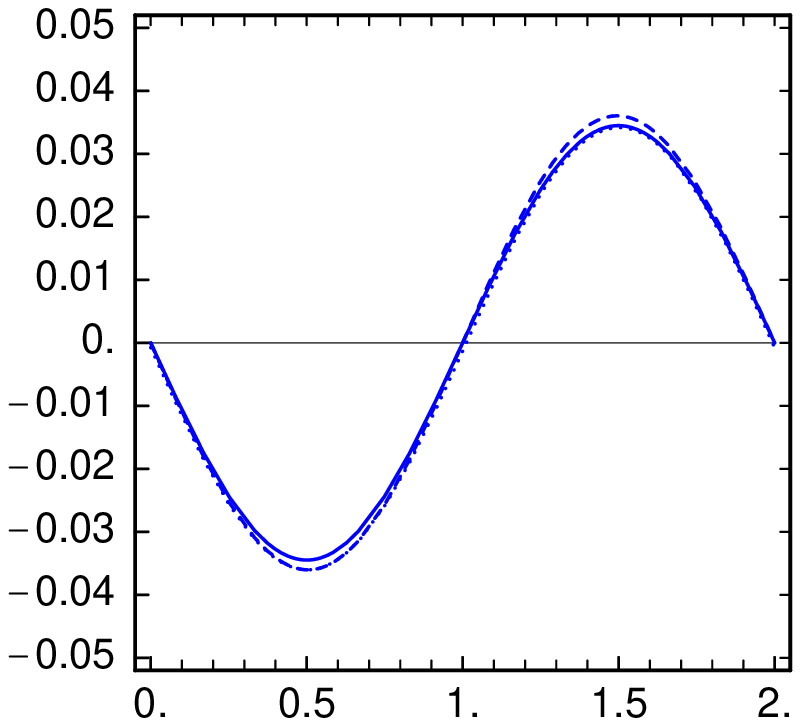,height=63mm}}}
\put(40,13){\footnotesize\mbf $\tan\b=40$}
\put(32,-2){$\phi_t~[\pi]$}
\put(-5,31){\rotatebox{90}{$\d^{CP}$}}
\put(-5,56){\bf b)}
\end{picture}
\caption{$\d^{CP}$ as a function of $\phi_t$, for $M_{\ti Q}=350$~GeV,
  $\mhp=700$~GeV, $\tan\b=10$ in {\bf (a)} and $\tan\b=40$ in {\bf (b)};
  full lines: $\phi_b=0$, dashed lines: $\phi_b=\phi_t$,
  dotted lines: $\mu=350$~GeV, $\phi_b=0$.
  The other parameters are fixed by Eq.~\eq{parset}.
  \label{fig:S2At}}
\end{center}
\end{figure}

The dependence on $\phi_t$ is shown explicitly in \fig{S2At},
where we plot $\d^{CP}$ as a function of $\phi_t$ for
$M_{\ti Q}=350$~GeV, $\mhp=700$~GeV, $\tan\b=10$ and 40 and
various choices of $\phi_b$.
As expected, $\d^{CP}$ shows a $\sim\sin\phi_t$ dependence.

%
%

\begin{figure}[h!]
\setlength{\unitlength}{1mm}
\begin{center}
\begin{picture}(60,64)
\put(0,0){\mbox{\epsfig{figure=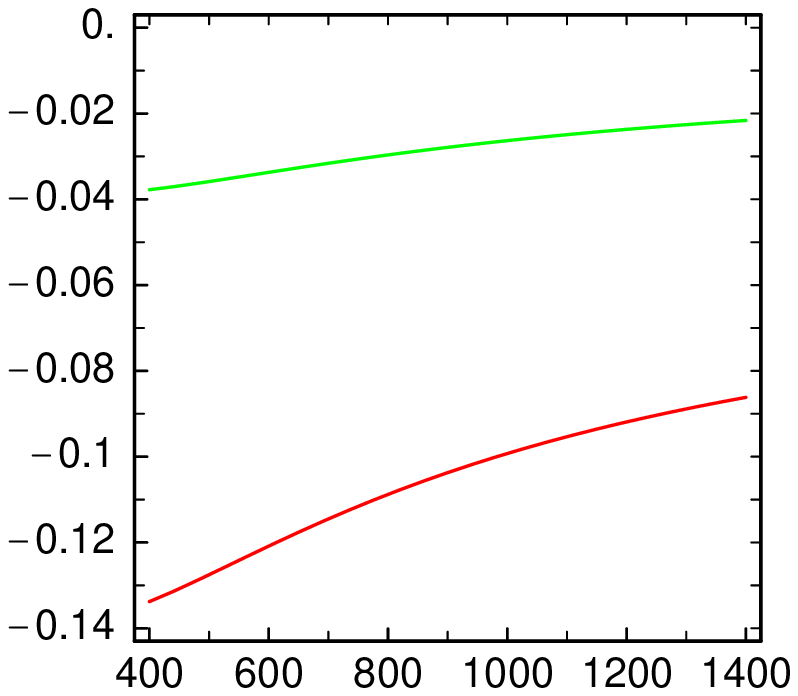,height=63mm}}}
\put(30,27){\footnotesize $\tan\b=10$}
\put(17,49){\footnotesize $\tan\b=40$}
\put(28,-1){$\msg$~[GeV]}
\put(-5,30){\rotatebox{90}{$\d^{CP}$}}
\end{picture}
\hspace{16mm}
\begin{picture}(60,62)
\put(0,0){\mbox{\epsfig{figure=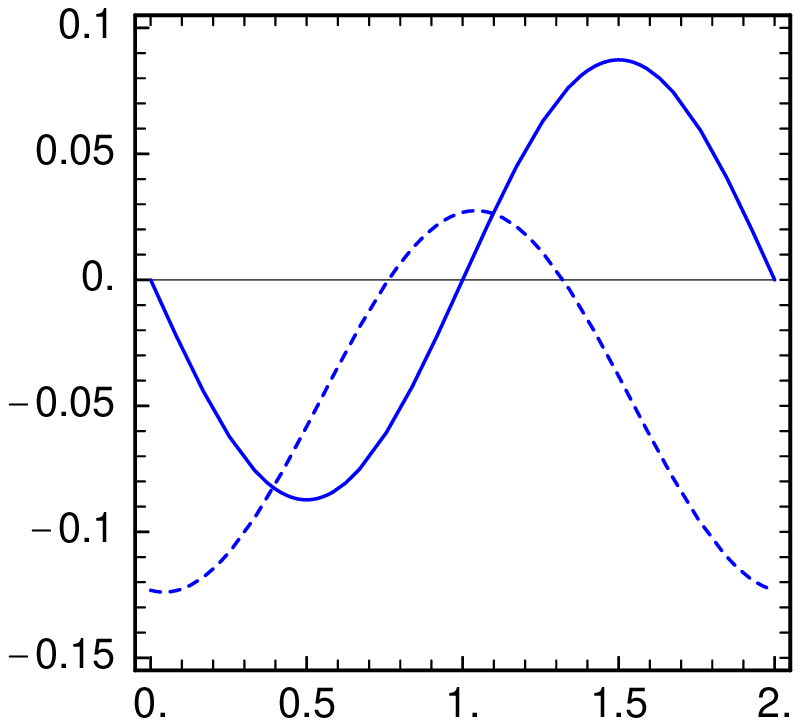,height=63mm}}}
\put(31,51){\footnotesize $\phi_t=0$}
\put(40,20){\footnotesize $\phi_t=\frac{\pi}{2}$}
\put(32,-2){$\phi_3~[\pi]$}
\put(-5,31){\rotatebox{90}{$\d^{CP}$}}
\put(-5,56){\bf b)}
\end{picture}
\caption{{\bf a)} $\d^{CP}$ as a function of $\msg$, for $M_{\ti Q}=350$~GeV,
  $\mhp=700$~GeV, $\phi_t=\pi/2$, and $\tan\b=10$ and 40.
  {\bf b)} $\d^{CP}$ as a function of $\phi_3$, the phase of $M_3$, 
  for $\msg=565$~GeV, $M_{\ti Q}=350$~GeV, $\mhp=700$~GeV, $\tan\b=10$, 
  $\phi_t=0$ (full line) and $\phi_t=\pi/2$ (dashed line).
  $\phi_b=0$, the other parameters are fixed by Eq.~\eq{parset}.
  \label{fig:S2msg}}
\end{center}
\end{figure}

Last but not least we relax the GUT relations between the gaugino 
masses and take $\msg$ as a free parameter (keeping, however, the 
relation between $M_1$ and $M_2$). 
\Fig{S2msg}a shows the dependence of $\d^{CP}$ on the gluino mass, 
for $M_{\ti Q}=350$~GeV, $\mhp=700$~GeV, $\phi_t=\pi/2$, and
$\tan\b=10$ and 40 (keeping $M_3=\msg$ real). 
It is interesting that there is still an effect for a large gluino mass: 
for $\tan\b=10$, $\d^{CP}$ is reduced from about $-12\%$ to
about $-9\%$ for $\msg=600\to 1200$~GeV. 
Also in the large $\tan\b$ case, $\d^{CP}$ is decreased by 30\% 
when the gluino mass is doubled: 
from $-3.4\%$ to $-2.4\%$ for $\msg=600\to 1200$~GeV.
A non-zero phase of $M_3$ may also have a large effect. 
This is shown in \fig{S2msg}b, where we plot $\d^{CP}$ 
as a function of $\phi_3$ for $\msg=|M_3|=565$~GeV and 
$\phi_t=0$ and $\pi/2$. 
In both cases, the asymmetry can be ${\cal O}(10\%)$. 
For $\mu>0$, the curves are shifted but the order of magnitude 
does not change. 

%
%

\section{Conclusions}

We have calculated the difference between the partial rates 
$\G\,(H^+ \to t\bar b)$ and $\G\,(H^- \to \bar tb)$ 
due to CP-violating phases in the MSSM. 
The resulting decay rate asymmetry $\d^{CP}$, Eq.~(1), 
could be measured in a counting experiment. 
If $\mhp < \mst{1}+\msb{1}$, $\d^{CP}$ is typically 
of the order of $10^{-3}$. 
However, for $\mhp > \mst{1}+\msb{1}$, $\d^{CP}$ can go up 
to 10\,--\,15\%, depending on the phases of $A_t$ and $A_b$, 
and on $\tan\b$. Also a phase of $M_3$ can have a large effect.

At the Tevatron, no sensitivity for detecting $H^\pm$ is 
expected for a mass $\mhp\gsim 200$~GeV. 
The LHC, on the other hand, has a discovery reach up to 
$\mhp\sim 1$~TeV, especially if QCD and SUSY effects conspire 
to enhance the cross section. With a luminosity of ${\cal L}=100$~fb$^{-1}$, 
about 217 signal events can be expected for $pp\to H^+\bar t b$ with 
$S/\sqrt{B}=6.3$ ($B$ being the background), 
for $\mhp\simeq 700$~GeV and $\tan\b=50$~\cite{Belyaev:2002eq}. 
In $e^+e^-$ collisions, the dominant production mode is 
$e^+e^-\to H^+H^-$. Therefore, one would need a centre-of-mass 
energy $\sqrt{s}>2\mhp$. This would certainly be realized 
at a multi-TeV linear collider such as CLIC~\cite{Battaglia:2001be}.
Hence, a CP-violating asymmetry $\d^{CP}$ of a few per cent 
should be measurable at the LHC or CLIC.

\section*{Acknowledgements}

We thank Johann K\"uhn and Werner Porod for helpful discussions.
We also thank Marco Fabbrichesi for his participation 
in the initial stage of this work.
The work of E.\,C. was supported by the Bulgarian National Science 
Foundation, Grant Ph--1010.

\begin{appendix}

\section{Masses and mixing matrices}

%
%

The {\bf neutralino} mass matrix in the basis of
\begin{equation}
  \Psi_j^0=\left(-i\lambda ',-i\lambda^3,\psi_{H_1}^0,\psi_{H_2}^0\right)
\end{equation}
is:
\begin{equation}
  {\cal M}_N =
  \left( \begin{array}{cccc}
  M_1 & 0 & -m_Z\sin\theta_W\cos\beta  & m_Z\sin\theta_W\sin\beta \\
  0 & M_2 &  m_Z\cos\theta_W\cos\beta  & -m_Z\cos\theta_W\sin\beta  \\
  -m_Z\sin\theta_W\cos\beta & m_Z\cos\theta_W\cos\beta   & 0 & -\mu \\
   m_Z\sin\theta_W\sin\beta & - m_Z\cos\theta_W\sin\beta & -\mu & 0
  \end{array}\right)
\end{equation}
with $\tan\beta = v_2/v_1$. This matrix is diagonalized by the
unitary mixing matrix $N$:
\begin{equation}
  N^*{{\cal M}_N} N^\dagger =
  {\rm diag}(\mnt{1},\,\mnt{2},\,\mnt{3},\,\mnt{4})\,,
\end{equation}
where $\mnt{k}$, $k=1,...,4$, are the (non-negative) masses
of the physical neutralino states.

%
%

\noi
The {\bf chargino} mass matrix is:
\begin{equation}
  {\cal M}_C =
  \left( \begin{array}{cc}
    M_2 &\sqrt 2\, m_W\sin\beta \\
    \sqrt 2\,m_W\cos\beta & \mu
  \end{array}\right) \,.
\end{equation}
It is diagonalized by the two unitary matrices $U$ and $V$:
\begin{equation}
  U^*{\cal M}_C V^\dagger = {\rm diag}(\mch{1},\,\mch{2})\,,
\end{equation}
where $\mch{1,2}$ are the masses of the physical chargino states.

%
%

\noi
The mass matrix of the {\bf stops}
in the basis $(\tilde t_L,\,\tilde t_R)$ is
\begin{equation}
  {\cal M}_{\tilde t}^2 = \left(
  \begin{array}{cc}
    M_{\ti Q}^2 + m_Z^2\cos 2\beta(\sfrac{1}{2}-\sfrac{2}{3}\sin^2\theta_W)
    + m_t^2 & (A_t^*-\mu\cot\beta)\,m_t \\
    (A_t-\mu^*\cot\beta)\,m_t
    & M_{\ti U}^2 + \sfrac{2}{3}m_Z^2\cos 2\beta\sin^2\theta_W + m_t^2
  \end{array}\right) \,.
\end{equation}
${\cal M}_{\tilde t}^2$ is diagonalized by the rotation matrix
$R^{\,\st}$ such that
$R^{\,\st\,\dagger}\,{\cal M}_{\tilde t}^2\, R^{\,\st} =
 {\rm diag}(m_{\tilde t_1}^2,\,m_{\tilde t_2}^2)$ and
${\scriptsize \Big(\!\!\begin{array}{cc} \st_L \\ \st_R \end{array}\!\!\Big)} = R^{\tilde t} \,
 {\scriptsize \Big(\!\!\begin{array}{cc} \st_1 \\ \st_2 \end{array}\!\!\Big)}$.
We have:
\begin{equation}
  R^{\,\tilde{t}} =
    \left(\begin{array}{rr}
      R^{\,\tilde{t}}_{L1} & R^{\,\tilde{t}}_{L2} \\
      R^{\,\tilde{t}}_{R1} & R^{\,\tilde{t}}_{R2}
    \end{array}\right)
  = \left(\begin{array}{rr}
      e^{\frac{i}{2} \varphi_{\tilde{t}}} \cos\theta_{\tilde{t}}
    & -e^{\frac{i}{2} \varphi_{\tilde{t}}} \sin\theta_{\tilde{t}}
  \\ e^{-\frac{i}{2} \varphi_{\tilde{t}}} \sin\theta_{\tilde{t}}
    & e^{-\frac{i}{2} \varphi_{\tilde{t}}} \cos\theta_{\tilde{t}}
    \end{array}\right)
\; .
\end{equation}

\noi
Analogously, the mass matrix of the {\bf sbottoms}
in the basis $(\tilde b_L,\,\tilde b_R)$,
\begin{equation}
  {\cal M}_{\sb}^2 = \left(
  \begin{array}{cc}
    M_{\ti Q}^2 - m_Z^2\cos 2\beta(\sfrac{1}{2}-\sfrac{1}{3}\sin^2\theta_W)
    + m_b^2 & (A_b^*-\mu\tan\beta)\,m_b \\
    (A_b-\mu^*\tan\beta)\,m_b
    & M_{\ti D}^2 - \sfrac{1}{3}m_Z^2\cos 2\beta\sin^2\theta_W + m_b^2
  \end{array}\right) \,,
\end{equation}
is diagonalized by the rotation matrix $R^{\,\sb}$ such that
$R^{\,\sb\,\dagger}{\cal M}_{\sb}^2\, R^{\,\sb} =
 {\rm diag}(\msb{1}^2,\,\msb{2}^2)$.

%
%
In the {\bf neutral Higgs sector}, we have two CP-even states  
$\phi_i = \sqrt{2}\,\Re ({\cal H}_i^i) - v_i$, $i = 1,2$, and
and one CP-odd state 
$a=\sqrt{2}\,(-\sin\beta\,\Im({\cal H}_1^1)+\cos\beta\,\Im({\cal H}_2^2))$,
where ${\cal H}_1$ and ${\cal H}_2$ are the two Higgs doublets in the 
interaction basis. In the basis $(\phi_1, \phi_2, a)$,
the neutral Higgs mass matrix ${\cal M}_H^2$ can be written as
the well-known tree-level part, which has a block form in this basis,
plus a general $3 \times 3$ matrix containing the loop corrections:
\begin{equation}
{\cal M}_H^2 =
  \left(\begin{array}{ccc}
    {\rm s}^2\beta\,m_A^2 + {\rm c}^2\beta\,m_Z^2
    & -{\rm s}\beta\,{\rm c}\beta\,(m_A^2 + m_Z^2) & 0 \\
    -{\rm s}\beta\,{\rm c}\beta\,(m_A^2 + m_Z^2)
    & {\rm c}^2\beta\,m_A^2 + {\rm s}^2\beta\,m_Z^2 & 0 \\
    0 & 0 & m_A^2
  \end{array} \right) + \, \left({\cal M}_H^{\rm loop}\right)^2\,,
  \label{eq:M2higgsneutraltree}
\end{equation}
where ${\rm s}\beta\equiv\sin\b$, ${\rm c}\beta\equiv\cos\b$, etc.
In the case of complex parameters, the loop contributions of
$({\cal M}_H^{\rm loop})^2$ lead to a mixing of the CP-even
and CP-odd states. The mass eigenstates then are
\begin{equation}
  \left(\begin{array}{c} H_1^0\\ H_2^0\\ H_3^0 \end{array}\right) =
  \, O^T .\,
  \left(\begin{array}{c} \phi_1\\ \phi_2\\ a \end{array}\right) \,.
  \label{eq:higgsEVP}
\end{equation}
The real $3 \times 3$ rotation matrix $O$ diagonalizes
the mass matrix ${\cal M}_H^2$,
\begin{equation}
  O^T {\cal M}_H^2\,O =
  {\rm diag}\left(m^2_{H_1^0}, m^2_{H_2^0},m^2_{H_3^0}\right)\, ,
  \label{eq:higgsEVPeqn}
\end{equation}
with $m_{H_1^0} < m_{H_2^0} < m_{H_3^0}$.
The transformations of the Higgs fields from the interaction basis
to the mass eigenstate basis are given by
\begin{eqnarray}
  {\cal H}_1^1 & = & v_1 + \frac{1}{\sqrt{2}}\left[
      \left(O_{1j} + i\,\sin\beta O_{3j}\right)\,H_j^0
      -i\,\cos\beta \, G^0\right] \,,
  \nonumber\\
  {\cal H}_1^2 & = & -\cos\beta\, G^- + \sin\beta \, H^- \,,
  \nonumber\\
  {\cal H}_2^1 & = & \sin\beta \, G^+ + \cos\beta \, H^+ \,,
  \nonumber\\
  {\cal H}_2^2 & = & v_2 + \frac{1}{\sqrt{2}}
      \left[\left(O_{2j} + i\,\cos\beta O_{3j}\right)\,H_j^0
            + i\,\sin\beta \, G^0\right]\,,
   \label{eq:Higgstransf}
\end{eqnarray}
with the implicit sum over $j = 1,2,3$,
\begin{equation}
        v_1 = v\cos\beta = \frac{\sqrt{2}}{g}\,m_W\cos\beta\,,
  \qquad
        v_2 = v\sin\beta = \frac{\sqrt{2}}{g}\,m_W\sin\beta\,.
\label{eq:vevtransf}
\end{equation}
For the numerical evaluation of the physical Higgs masses
and the rotation matrix $O$ in the 1-loop effective potential
approach~\cite{CEPW00}, we use the program {\tt chp.f}~\cite{cph}.

\section{Interaction Lagrangian}

In this section we give the parts of the interaction Lagrangian
that we need for our calculation. We start with the interaction
of Higgs bosons with quarks and squarks:
\begin{equation}
{\cal L}_{Hqq} =
  H^+\,\bar{t}\,(y_b\PR + y_t\PL)\,b + H^-\,\bar{b}\,(y_t\PR + y_b\PL)\,t
  + H^0_l\,\bar q\,(s_l^{q,R}\PR + s_l^{q,L}\PL)\, q \,,
\end{equation}
\begin{equation}
{\cal L}_{H\sq\sq} =
  (G_4^{})_{ij}^{}\,H^+\,\st_i^*\,\sb_j^{} +
  (G_4^*)_{ij}^{}\, H^-\,\sb_j^*\,\st_i^{} \,,
\end{equation}
with $i,j=1,2$, $l=1,2,3$ and
\begin{equation}
   \PL = \smaf{1}{2}(1-\g_5)\,, \quad
   \PR = \smaf{1}{2}(1+\g_5)\,.
\end{equation}
\noi
For the $H^\pm$ couplings to quarks we have
\begin{equation}
  y_t = h_t\cos\b \,,\qquad
  y_b = h_b\sin\b \,,
\end{equation}
with
\begin{equation}
  h_t = \frac{g\,\overline{m}_t}{\sqrt 2\,m_W \sin\beta}\,, \qquad
  h_b = \frac{g\,\overline{m}_b}{\sqrt 2\,m_W\cos\beta}\,,
\end{equation}
where $\overline{m}_q$ is the $\overline{DR}$ running quark mass.
%
For the $H^0_l$ couplings to quarks we have
\begin{eqnarray}
  s_l^{q,R} & = &  - \frac{g\,\overline{m}_q}{2\, m_W}\,
    \left(g_{H_l qq}^S + i\, g_{H_l qq}^P\right) \,,
    \label{eq:slbRdef}\\
  s_l^{q,L} & = &  - \frac{g\,\overline{m}_q}{2\, m_W}\,
    \left(g_{H_l qq}^S - i\, g_{H_l qq}^P\right) \,,
  \label{eq:slbLdef}
\end{eqnarray}
with $g_{H_l qq}^S$ and $g_{H_l qq}^P$ given by Eqs.~(4.11)--(4.14)
in~\cite{CEPW00}.
%
The $H^\pm$ couplings to squarks are given by the matrix
\begin{equation}
   G_4^{} = R^{\,\st\,\dagger} \; \hat G_4^{} \; R^{\,\sb}\,,
\label{eq:G4sq}
\end{equation}
with
\begin{equation}
\hat G_4^{ } =
  \left(\! \begin{array}{cc}
    h_b m_b \sin\b + h_t m_t\cos\b - \rzw\,g\,m_W\sin\b\cos\b
    & h_b\,(A_b^*\sin\b + \mu\cos\b) \\[3mm]
      h_t\,(A_t\cos\b + \mu^*\sin\b) & h_t m_b \cos\b + h_b m_t \sin\b
\end{array}\! \right) \,.
\label{eq:GLR4}
\end{equation}
\vspace*{1mm}

\noi
The interactions of charginos and neutralinos are described by
\begin{eqnarray}
{\cal L}_{H\ti\x\ti\x}
  &=& H^+\bar{\ti\x}^+_i (F_{ik}^R\PR + F_{ik}^L\PL)\,\nt_k +
      H^-\bar{\ti\x}^0_k (F_{ik}^{R*}\PL + F_{ik}^{L*}\PR)\,\chp_i \,,\\
  {\cal L}_{q\sq\ti\x^+}
  &=& \bar t\,(l_{ij}^{\sb}\PR + k_{ij}^{\sb}\PL)\,\ti\x^{+}_j\,\sb_i^{} +
      \bar b\,(l_{ij}^{\st}\PR + k_{ij}^{\st}\PL)\,\ti\x^{+c}_j\,\st_i^{}
      \nn \\
  & &
  +\,\overline{\ti\x^{+}_j}\,(l_{ij}^{\sb*}\PL+k_{ij}^{\sb*}\PR)\,t\,\sb_i^*
  +\overline{\ti\x^{+c}_j}\,(l_{ij}^{\st*}\PL+k_{ij}^{\st*}\PR)\,b\,\st_i^*
  \,,\\
  {\cal L}_{q\sq\ti\x^0}
  &=& \bar q\,(a_{ik}^{\sq}\PR + b_{ik}^{\sq}\PL)\,\nt_k\,\sq_i^{}
   +\bar{\ti\x}^0_k (a_{ik}^{\sq*}\PL + b_{ik}^{\sq*}\PR)\,q\,\sq_i^*\,,
\end{eqnarray}

\noi
with $i,j=1,2$ and $k=1,...,4$.
The couplings of $H^\pm$ to charginos and neutralinos are
\begin{eqnarray}
  F_{ik}^R &=& -g\,\left[ V_{i1}N_{k4}
    + \smaf{1}{\rzw}\,(N_{k2} + N_{k1}\tan\t_W)\,V_{i2}
    \,\right]\,\cos\b\,,\\
  F_{ik}^L &=& -g\,\left[ U_{i1}^* N_{k3}^*
    - \smaf{1}{\rzw}\,(N_{k2}^*+N_{k1}^*\tan\t_W)\,U_{i2}^*
    \,\right]\,\sin\b\,.
\end{eqnarray}

\noi
The chargino--squark--quark couplings are
\begin{align}
  l_{ij}^{\st} &= -g\,V_{j1}\Rst_{1i} + h_t\,V_{j2}\Rst_{2i}\,, &
  l_{ij}^{\sb} &= -g\,U_{j1}\Rsb_{1i} + h_b\,U_{j2}\Rsb_{2i}\,, \\
  k_{ij}^{\st} &= h_b\,U_{j2}^*\Rst_{1i} \,, &
  k_{ij}^{\sb} &= h_t\,V_{j2}^*\Rsb_{1i} \,.
\end{align}

\noi
The neutralino--squark--quark couplings are
\begin{eqnarray}
  a_{ik}^{\,\sq} &=& g f_{Lk}^{\sq} \Rsq_{1i}
                     + h_{Lk}^{\sq*}\Rsq_{2i} \,,\\
  b_{ik}^{\,\sq} &=&   h_{Lk}^{\sq} \Rsq_{1i}
                   + g f_{Rk}^{\sq} \Rsq_{2i} \,,
\end{eqnarray}
with
\begin{align}
  f_{Lk}^{\st} &= -\sfrac{1}{\sqrt 2}\,(N_{k2}
                        +\sfrac{1}{3}\tan\theta_W N_{k1}) \,, &
  f_{Lk}^{\sb} &= \sfrac{1}{\sqrt 2}\,(N_{k2}
                        -\sfrac{1}{3}\tan\theta_W N_{k1}) \,,\\
  f_{Rk}^{\st} &= \sfrac{2\sqrt 2}{3}\,\tan\theta_W N_{k1}^*\,, &
  f_{Rk}^{\sb} &= -\sfrac{\sqrt 2}{3}\,\tan\theta_W N_{k1}^*\,,\\
  h_{Lk}^{\st} &= -h_t\, N_{k4}^* \,, &
  h_{Lk}^{\sb} &= -h_b\, N_{k3}^* \,.
\end{align}

\noi
Finally, the squark--quark--gluino interaction is given by
\begin{eqnarray}
  {\cal L}_{q\sq\sg} &=&
  -\rzw\,g_s\,T_{st}^a\left[\,
      \bar{\sg}^a (R_{1i}^{\sq}\,e^{-\frac{i}{2}\phi_3}\PL -
                   R_{2i}^{\sq}\,e^{ \frac{i}{2}\phi_3}\PR)\,
      q_s^{}\,\sq_{i,t}^* \right.\nn\\
  & & \hspace{3cm}\left.
      +\,\bar q_s^{} (R_{1i}^{\sq*}\,e^{ \frac{i}{2}\phi_3}\PR -
                   R_{2i}^{\sq*}\,e^{-\frac{i}{2}\phi_3}\PL)\,\sg^a\,\sq_{i,t}
      \,\right]\,,
\end{eqnarray}
and the quark interaction with $W$ bosons is
\begin{equation}
  {\cal L}_{qqW} =
  -\smaf{g}{\rzw}\,( W_\mu^+ \bar t\,\g^\mu\PL\,b +
                     W_\mu^- \bar b\,\g^\mu\PL\,t ) \,.
\end{equation}

\noi
We next turn to the interaction of Higgs bosons with $W$ bosons
and ghosts. The Lagrangian of two Higgs particles and one $W$
boson is given by
\begin{equation}
  \hspace*{-4mm}
  {\cal L}_{HHW} = -i\,\frac{g}{\sqrt{2}}\, \left[
    W_\mu^+ \Big(\H_1^{1*} \lrd \H_1^2 + \H_2^{1*} \lrd \H_2^2\Big)
  + W_\mu^- \Big(\H_1^{1} \lrd \H_1^{2*} + \H_2^{1} \lrd \H_2^{2*}\Big)
  \right]\,,
  \label{eq:lagHHV}
\end{equation}
where
\begin{equation}
  A\delr B = A\,(\partial_\mu B) - (\partial_\mu A)\,B \,.
\end{equation}
Using the transformations Eq.~(\ref{eq:Higgstransf}) we get
\begin{equation}
  \hspace*{-4mm}
  {\cal L}_{HHW} = i\, \frac{g}{2}\, \left[ W_\mu^+ \Big(
      g_{H_j H^- W^+} H^0_j \lrd H^-
    + g_{H_j G^- W^+} H^0_j \lrd G^-
    + i\, G^0 \lrd G^- \Big) + {\rm h.\,c.}\right]\,
  \label{eq:lagWH0Hj}
\end{equation}
where
\begin{eqnarray}
  g_{H_j H^- W^+} &=& -\sin\beta\,O_{1j} + \cos\beta\,O_{2j} + i\,O_{3j}
  \quad {\rm and} \\
  g_{H_j G^- W^+} &=& \cos\beta\, O_{1j} + \sin\beta\, O_{2j} \,.
  \label{eq:gH0HmWp}
\end{eqnarray}
Moreover, $g_{H_j H^+ W^-} = g^*_{H_j H^- W^+}$ and
$g_{H_j G^+ W^-} = g_{H_j G^- W^+}$.
Note that there is no $G^0\,W^\pm\,H^\mp$ coupling.
We further need the couplings of two charged and one neutral Higgs bosons.
They are derived from the $D$-term interaction Lagrangian of the 
Higgs sector,\\
${\cal L} = - \frac{1}{2}\left(D' D' + D^1 D^1 + D^2 D^2 + D^3 D^3\right)$,
where $D'$ and $D^i$ are the $U(1)_Y$ and  $SU(2)_I$ $D$-terms, respectively.
In terms of Higgs fields in the interaction basis we have
\begin{eqnarray}
  {\cal L}_{HHH} &=&  -\frac 1 8 \left(g^2 + {g'}^2\right)
  \left(\H_1^{1*} \H_1^1 + \H_1^{2*} \H_1^2 -  \H_2^{1*} \H_2^1
        - \H_2^{2*} \H_2^2\right) \nonumber\\
  && \hspace{2.6cm}
  -\,\frac{g^2}{2} \left( \H_1^{1*} \H_2^1 + \H_1^{2*} \H_2^2\right)
  \left( \H_1^1 \H_2^{1*} + \H_1^2 \H_2^{2*}\right)\, .
  \label{eq:lagHHHH}
\end{eqnarray}
The Lagrangian in the mass eigenstate basis is again obtained by
applying the transformations of Eqs.~(\ref{eq:Higgstransf}) and
(\ref{eq:vevtransf}). We are only interested in the combinations
of $(H^0_l,\, G^0) \times (H^\pm,\, G^\pm) \times (H^\mp,\, G^\mp)$.
The couplings to $G^0$, e.g. $G^0 H^+ H^-$, $G^0 H^+ G^-$, $G^0 G^+ H^-$,
and $G^0 G^+ G^-$, are zero.
The couplings to $H^0_l$ are:
\begin{eqnarray}
   {\cal L}_{HHH} &=& g_{H_j H^+ H^-}\, H^0_j H^+ H^-
                    + g_{H_j H^+ G^-}\, H^0_j H^+ G^-  \nn\\
   & & \hspace{4cm} + g^*_{H_j H^+ G^-}\, H^0_j G^+ H^-
                    + g_{H_j G^+ G^-}\, H^0_j G^+ G^- \,,
  \label{eq:lagHHH}
\end{eqnarray}
with
\begin{eqnarray}
  g_{H_j H^+ H^-} &=& \frac{g\,m_W}{2}\,
    \left\{
    [(1+{\rm t}_W^2)\,{\rm c}2\beta - 2]\,{\rm c}\beta\,O_{1j} -
    [(1+{\rm t}_W^2)\,{\rm c}2\beta + 2]\,{\rm s}\beta\,O_{2j}
    \right\}\,,
    \label{eq:lagHjH+H-}\\
  g_{H_j H^+ G^-} &=& \frac{g\,m_W}{2}\,
    \left[\,
    {\rm c}2\beta\,({\rm s}\beta\,O_{1j}+{\rm c}\beta\,O_{2j}) -i\,O_{3j}
    + {\rm t}_W^2\,{\rm s}2\beta\,({\rm c}\,\beta O_{1j}
      - {\rm s}\beta\,O_{2j})
    \right] \, ,
    \label{eq:lagHjH+G-}\\
  g_{H_j G^+ G^-} &=&  - \frac{g\,m_W}{2}\,
   (1 + {\rm t}_W^2)\,{\rm c}2\beta\,
   ({\rm c}\beta\,O_{1j} - {\rm s}\beta\,O_{2j}) \,.
   \label{eq:lagHjG+G-}
\end{eqnarray}
Note that only $g_{H_j H^+ G^-}$ is complex.
In Eqs.~\eq{lagHjH+H-}\,--\,\eq{lagHjG+G-}, we have used the abbreviations
${\rm s}\beta\equiv\sin\b$, ${\rm s}2\beta\equiv\sin2\b$,
${\rm c}\beta\equiv\cos\b$, ${\rm c}2\beta\equiv\cos2\b$, and
${\rm t}_W^2 \equiv \tan^2\t_W$.

\section{Passarino--Veltman integrals \label{app:CX}}

Here we give the definition of the Passarino--Veltman
one-, two-, and three-point functions~\cite{pave} in the convention 
of \cite{Denner}. 
For the general denominators we use the notation
\begin{equation}
  {\mathcal D}^{0} = q^{2} - m_{0}^{2}
  \quad \mbox{and}\quad
  {\mathcal D}^{j} = ( q + p_{j} )^{2} - m_{j}^{2}\,.
\end{equation}
Then the loop integrals in $D=4-\epsilon$ dimensions are as follows:
\begin{eqnarray}
  A_0(m_0^2) &=& \frac{1}{i\pi^2}\int d^{D}\!q\:\frac{1}{{\mathcal D}^0}\,,\\
  B_0(p_1^2,m_0^2,m_1^2)   &=& \frac{1}{i\pi^2} \int d^{D}\! q \:
          \frac{1}{{\mathcal D}^0 {\mathcal D}^1} \,,\\
  B_\mu(p_1^2,m_0^2,m_1^2) &=& \frac{1}{i\pi^2} \int d^{D}\! q \:
      \frac{q_\mu}{{\mathcal D}^0 {\mathcal D}^1}
      = p_{1\mu}\, B_1(p_1^2,m_0^2,m_1^2) \,,
\end{eqnarray}
and
\begin{eqnarray}
  C_0 &=& \frac{1}{i\pi^2} \int d^{D}\! q \:
          \frac{1}{{\mathcal D}^0 {\mathcal D}^1 {\mathcal D}^2} \,,\\
  C_\mu &=& \frac{1}{i\pi^2} \int d^{D}\! q \:
      \frac{q_\mu}{{\mathcal D}^0 {\mathcal D}^1 {\mathcal D}^2}
      = p_{1\mu} C_1 + p_{2\mu} C_2  \,,\\
  C_{\mu\nu} &=& \frac{1}{i\pi^2} \int d^{D}\! q \:
      \frac{q_\mu q_\nu}{{\mathcal D}^0 {\mathcal D}^1 {\mathcal D}^2} \nn\\
      &=& g_{\mu\nu} C_{00} + p_{1\mu} p_{1\nu} C_{11}
          + ( p_{1\mu} p_{2\nu} + p_{2\mu} p_{1\nu} ) C_{12}
          + p_{2\mu} p_{2\nu} C_{22} \,,
\end{eqnarray}
where the $C$'s have
$(p_{1}^{2},(p_{1}-p_{2})^{2},p_{2}^{2},m_{0}^{2},m_{1}^{2},m_{2}^{2})$
as their arguments.
The function $B_1$ can be expressed as a combination of the 
functions $A_0$ and $B_0$:
\begin{equation}
  2p_1^2 B_1(p_1^2,m_0^2,m_1^2) = 
  A_0(m_0^2) - A_0(m_1^2) - (p_1^2-m_1^2+m_0^2)\,B_0(p_1^2,m_0^2,m_1^2)\,.
\end{equation}

\end{appendix}



\begin{thebibliography}{99}

\bibitem{Dugan:1984qf} M.~Dugan, B.~Grinstein and L.~J.~Hall,
                       Nucl.\ Phys.\ B {\bf 255} (1985) 413.


\bibitem{Carena:1997ki} M.~Carena, M.~Quiros and C.~E.~Wagner,
                        Nucl.\ Phys.\ B {\bf 524} (1998) 3 [hep-ph/9710401];
for a review see: A.~G.~Cohen, D.~B.~Kaplan and A.~E.~Nelson,
                  Ann.\ Rev.\ Nucl.\ Part.\ Sci.\  {\bf 43} (1993) 27
                  [hep-ph/9302210].


\bibitem{Altarev:cf} 
I.~S.~Altarev {\it et al.},
     Phys.\ Lett.\ B {\bf 276} (1992) 242;
I.~S.~Altarev {\it et al.}, 
     Phys.\ Atom.\ Nucl.\  {\bf 59} (1996) 1152
     [Yad.\ Fiz.\  {\bf 59N7} (1996) 1204];
E.~D.~Commins, S.~B.~Ross, D.~DeMille and B.~C.~Regan,
     Phys.\ Rev.\ A {\bf 50} (1994) 2960.


\bibitem{Nath:dn} 
P.~Nath, Phys.\ Rev.\ Lett.\ {\bf 66} (1991) 2565;
Y.~Kizukuri and N.~Oshimo, 
     Phys.\ Rev.\ D {\bf 46} (1992) 3025;
R.~Garisto and J.~D.~Wells, 
     Phys.\ Rev.\ D {\bf 55} (1997) 1611 [hep-ph/9609511];
Y.~Grossman, Y.~Nir and R.~Rattazzi,
     Adv.\ Ser.\ Direct.\ High Energy Phys.\ {\bf 15} (1998) 755
     [hep-ph/9701231].


\bibitem{Ibrahim:1997nc} 
T.~Ibrahim and P.~Nath,
     Phys.\ Lett.\ B {\bf 418} (1998) 98 [hep-ph/9707409];
M.~Brhlik, G.~J.~Good and G.~L.~Kane,
     Phys.\ Rev.\ D {\bf 59} (1999) 115004 [hep-ph/9810457];
A.~Bartl, T.~Gajdosik, W.~Porod, P.~Stockinger and H.~Stremnitzer,
     Phys.\ Rev.\ D {\bf 60} (1999) 073003 [hep-ph/9903402].


\bibitem{Atwood:2000tu} For a review, see:
                        D.~Atwood, S.~Bar-Shalom, G.~Eilam and A.~Soni,
                        Phys.\ Rept.\  {\bf 347} (2001) 1
                        [hep-ph/0006032].


\bibitem{Pilaftsis:1998pe} 
A.~Pilaftsis, 
     Phys.\ Rev.\ D {\bf 58} (1998) 096010 [hep-ph/9803297] and 
     Phys.\ Lett.\ B {\bf 435} (1998) 88 [hep-ph/9805373];
A.~Pilaftsis and C.~E.~Wagner,
     Nucl.\ Phys.\ B {\bf 553} (1999) 3 [hep-ph/9902371];
D.~A.~Demir,
     Phys.\ Rev.\ D {\bf 60} (1999) 055006 [hep-ph/9901389].


\bibitem{CEPW00} M.~Carena, J.~R.~Ellis, A.~Pilaftsis and C.~E.~Wagner,
                 Nucl.\ Phys.\ B {\bf 586} (2000) 92 [hep-ph/0003180].


\bibitem{pave} G.~Passarino and M.~J.~Veltman,
               Nucl.\ Phys.\ B {\bf 160} (1979) 151.


\bibitem{Denner}
A.~Denner, Fortschr. Phys. {\bf 41} (1993) 307.


\bibitem{SPS} 
B.~C.~Allanach {\it et al.}, hep-ph/0202233, 
     in {\it Proc.\ of the APS/DPF/DPB Summer Study on the Future of 
     Particle Physics (Snowmass 2001) }, eds.\ R.~Davidson and C.~Quigg.


\bibitem{cph} The Fortran program {\tt cph.f} can be obtained from \\
              {\tt http://pilaftsi.home.cern.ch/pilaftsi/}


\bibitem{published}
E. Christova, H. Eberl, S. Kraml and W. Majerotto, 
     Nucl.\ Phys.\ B {\bf 639} (2002) 263.

\bibitem{erratum}
E. Christova, H. Eberl, S. Kraml and W. Majerotto, 
     Erratum to Nucl.\ Phys.\ B {\bf 639} (2002) 263, 
     to appear.


\bibitem{Carena:2001fw} 
M.~Carena, J.~R.~Ellis, A.~Pilaftsis and C.~E.~Wagner,
     Nucl.\ Phys.\ B {\bf 625} (2002) 345 [hep-ph/0111245];
S.~Heinemeyer,
     Eur.\ Phys.\ J.\ C {\bf 22} (2001) 521 [hep-ph/0108059];
T.~Ibrahim and P.~Nath,
     hep-ph/0204092.


\bibitem{Belyaev:2002eq} 
A.~Belyaev, D.~Garcia, J.~Guasch and J.~Sola, hep-ph/0203031.


\bibitem{Battaglia:2001be} 
M.~Battaglia, A.~Ferrari, A.~Kiiskinen and T.~Maki, hep-ex/0112015, 
     in {\it Proc.\ of the APS/DPF/DPB Summer Study on the Future of 
     Particle Physics (Snowmass 2001) }, eds.\ R.~Davidson and C.~Quigg.

\end{thebibliography}
\end{document}